\documentclass[12pt]{iopart}

\usepackage{iopams}
\usepackage{amssymb}
\usepackage{setstack}
\usepackage{graphicx}
\usepackage{dcolumn}
\usepackage{bm}
\usepackage{dsfont}
\usepackage{amssymb}

\newcommand{\atanh  }{{\rm{atanh}}}
\newcommand{\atan   }{{\rm{arctan}}}

\newcommand{\bM     }{\mbox{\boldmath$M$}}

\newcommand{\bQ     }{\mbox{\boldmath$Q$}}

\newcommand{\bC     }{\mbox{\boldmath$C$}}

\newcommand{\bA     }{\mbox{\boldmath$A$}}

\newcommand{\s      }{{\rm sign}}

\usepackage{color}

\begin{document}
\date{\today}
\title{\bf The Cavity Approach to Parallel Dynamics of Ising Spins on a Graph}
\author{I. Neri and D. Boll\'e}
\address{Instituut voor Theoretische Fysica, Katholieke Universiteit Leuven, Celestijnenlaan 200D, B-3001 Leuven, Belgium}
\ead{izaak.neri@fys.kuleuven.be, desire.bolle@fys.kuleuven.be}
\begin{abstract}
We use the cavity method to study parallel dynamics of disordered Ising models on a graph.  In particular, we derive a set of recursive equations in single site probabilities of paths propagating along the edges of the graph.  These equations are analogous to the cavity equations for equilibrium models and are exact on a tree.  

On graphs with exclusively directed edges we find an exact expression for the stationary distribution of the spins.  We present the phase diagrams for an Ising model on an asymmetric Bethe lattice and for a neural network with Hebbian interactions on an asymmetric scale-free graph.

  For graphs with a nonzero fraction of symmetric edges the equations can be solved for a finite number of time steps.  Theoretical predictions are confirmed by simulation results.  

Using a heuristic method, the cavity equations are extended to a set of equations that determine the marginals of the stationary distribution of Ising models on graphs with a nonzero fraction of symmetric edges.  The results of this method are discussed and compared with simulations.  
 \end{abstract}
\section{Introduction}
Many problems in different research fields are based upon the interaction of units through some underlying graph.  Some examples are: gene expressions in boolean networks \cite{Corr2006}, agents competing for some limited resources \cite{Coolen2005, Paczus2000}, interactions between the decoding variables in low-density parity check codes \cite{Vic1999}, interactions between humans on a social network \cite{Klein07}, the analysis of phase transitions of a spin glass \cite{viana1985}.

 To calculate statistical quantities on a given graph instance, one can use the cavity method \cite{Mez2001}.  This method is based on the assumption that for sparse graphs the neighbouring spins only depend on each other through their direct interactions.  A similar method, known as the sum-product algorithm, is used in information theory and artificial intelligence, see \cite{Kschi2001} for a tutorial paper.  Examples of problems investigated with the cavity method are: the characterisation of the set of solutions of optimisation problems on random graphs \cite{Flor2008}, the calculation of the eigenvalue spectrum of sparse random matrices \cite{Tim2008} and the solution of the minimum weight Steiner tree problem \cite{Steiner}. 

For many problems the stationary distribution of the spins is not known, e.g., neural networks with asymmetric couplings \cite{Cris1988}, the minority game \cite{Coolen2005}, ... One would like to generalize the cavity method to treat the internal dynamics of these models.
One of the standard approaches here is the generating functional analysis first discussed in \cite{Dominicis1978}.  The stationary solution of the minority game was found using this method \cite{Coolen2005}.  In work in progress \cite{Mimura}  the parallel dynamics on graphs is studied using the generating functional method of \cite{Hat2005}.  The same authors used this method to analyze the evolution of a decoding algorithm on a sparse graph \cite{Mimura2}.  Another successful method is the dynamical replica analysis which was applied on finitely connected systems \cite{Hat2005, Mozeika2007}.  Recently, sequential dynamics of an Ising spin glass on a Bethe lattice with binary couplings was solved using a cavity-like approach \cite{Kiemes2008}. 

In this work we apply the cavity method to solve the parallel dynamics of models on random graphs.
The aim thereby is twofold.  First, we want to generalize the effective dynamical equations that solve the dynamics on a Poissonian graph, given in \cite{Hat2005}, to random graphs with a given degree distribution.  This generalisation is important when we want to solve the dynamics of models on a given graph structure.  We use these equations to solve the dynamics and to find the stationary distribution of an Ising model and a neural network model, both on a fully asymmetric graph.  We discuss how the correlations in the indegrees and outdegrees influence the performance of the neural network.  The second purpose of the paper is to extend the cavity equations to graphs with both symmetric and asymmetric couplings. For this we need to find the stationary state of the dynamics.  This is possible when we neglect the correlations in time of the stationary distribution.  We discuss how good this approach can predict macroscopic observables of Ising models with bond disorder or with fluctuating connectivities.    

This paper is organized as follows: In section \ref{sec:Dynamics} we define the necessary quantities and derive the effective dynamical equations for the single site marginals on a given graph instance.  We take the average of these over a graph ensemble in section \ref{sec:average}. In section
\ref{sec:bench} we specify the dynamics of the spin models.   We derive
the equations for the distributions of single site marginals in section \ref{sec:ensemble}.  The evolution of macroscopic observables obtained from the theory is compared  with simulations in section \ref{sec:simulation}. We discuss the phase diagrams for an Ising model on an asymmetric Bethe lattice and for a neural network with Hebbian interactions on a scale-free graph in sections \ref{sec:Bethe} and \ref{sec:Neural} respectively.  In section \ref{sec:9} we derive an algorithm that calculates the single site marginals of the stationary distribution on graphs with arbitrary symmetry.  A discussion is given in
section~\ref{sec:disc}.  

\section{Dynamics on a given graph instance}\label{sec:Dynamics}
\subsection{Some Definitions and Notations}
We consider models defined on a given graph instance $G = (V,E)$, with $V$ and $E$ respectively the set of vertices (or sites) and the set of edges.  We limit ourselves to simple directed graphs $G$ determined by a connectivity matrix $\bC$, with elements $\left[\bC\right]_{ij}=c_{ij} \in \left\{0,1\right\}$.  When $c_{ij}=1$ and $c_{ji}=0$ the graph has a directed edge from the $i$-th site to the $j$-th site.  When $c_{ij} = c_{ji} = 1$ there is an undirected edge between $i$ and $j$ and when $c_{ij}=c_{ji}=0$ there are no edges between them.  We define the sets $E$, $E^{\rm{d}}$ and $E^{\rm sym}$ through: $E \equiv\left\{(i,j)\in V\times V| c_{ij}=1\right\}$, $E^{\rm{d}} \equiv\left\{(i,j)\in V\times V| c_{ij}=1, c_{ji}=0\right\}$ and  $E^{\rm sym} \equiv\left\{(i,j)\in V\times V| c_{ij}=1, c_{ji}=1\right\}$.  We study the evolution of Ising like models of $n$-replicated Ising variables $\sigma_i(t) \in \left\{-1,1\right\}^n$, with $i=1, \ldots N$ and $t$ the corresponding discrete time step.   The dynamics in discrete time is defined by a transition probability $W(\bsigma(s)|\bsigma(s-1))$, from the state $\bsigma(s-1) = (\sigma_1(s-1), \sigma_2(s-1), \cdots, \sigma_N(s-1))$ on the $(s-1)$-th time step  to the state $\bsigma(s)$ on the $s$-th time step.  We consider transition probabilities $W$ of the form:
\begin{eqnarray}
\fl W\left[\bsigma(s)| \bsigma(s-1); \btheta\right] = \prod^N_{i=1}W\left[\sigma_i(s)|\bsigma(s-1); \theta_i(s)\right] =  \prod^N_{i=1}W\left[\sigma_i(s)|h_i(s)\right] \:.\label{eq:W}
\end{eqnarray}
The $n$-dimensional local field $h_i(s)$ is defined through 
\begin{eqnarray}
 h_i(s) = \sum_{j\in \partial^{\rm{in}}_i}h_{j\rightarrow i}(\sigma_j(s-1)) +\theta_i(s)\:, \label{eq:h}
\end{eqnarray}
where the field $h_{j\rightarrow i}(\sigma_j(s-1))$ quantifies the influence of the spin on site $j$ on the spin on site $i$ and $\theta_i(s)$ is an external field.  
We used $\partial^{\rm{in}}_i$ for the neighbourhood of all the vertices that influence $i$ directly, i.e. $\partial^{\rm{in}}_i \equiv \left\{j \in V|c_{ji}=1\right\}$.  We will also use: $\partial^{\rm{out}}_i \equiv \left\{j\in V|c_{ij}=1\right\}$, $\partial_i \equiv \partial^{\rm{in}}_i \cup \partial^{\rm{out}}_i$ and $\partial^{\rm sym}_i \equiv \partial^{\rm{in}}_i \cap\partial^{\rm{out}}_i$. The probability to have the path $\bsigma^{t_0..t} = \left(\bsigma(t_0), \cdots, \bsigma(t)\right)$, from time step $t_0$ to time step $t$, is given by
\begin{eqnarray}
 \fl P_{t_0..t}\left(\bsigma^{t_0..t}|\btheta^{t_0+1..t}\right)&= \left(\prod^t_{s=t_0+1}W\left[\bsigma(s)|\bsigma(s-1); \btheta(s)\right]\right)P_{t_0}\left(\bsigma(t_0)\right)\:,
\end{eqnarray}
with $P_{t_0}\left(\bsigma(t_0)\right)$ the probability distribution of the spins at time step $t_0$.

\subsection{Dynamical Version of the Cavity Equations}

Using the cavity method, see \cite{Mez2001}, it is possible to solve the parallel dynamics on graphs.  The cavity graph $G^{(i)}$ is the subgraph of $G$ where the $i$-th vertex and all of the interactions with its neighbours are removed.
We write the following relationship between a path probability $P_{0..t}$ on the graph $G$ and the probability $P^{(i)}_{0..t}$ on its related cavity graph $G^{(i)}$
\begin{eqnarray}
  \fl P_{0..t}\left(\bsigma^{0..t}|\btheta^{1..t}\right) =  P^{(i)}_{0..t}\left(\bsigma^{0..t}|\btheta^{1..t} + \bzeta^{(i), 1..t} \right)\left(\prod^t_{s=1}W\left[\sigma_i(s)|\bsigma(s-1); \theta_i(s)\right]\right)p_{0}(\sigma_i(0)) \:. \nonumber \label{eq:PathProb}\\  
\end{eqnarray}
In equation (\ref{eq:PathProb}) we introduced an extra field $\zeta^{(i)}_j(s)$, representing the influence of the $i$-th spin on its neighbours $j\in \partial^{\rm out}_i$:
\begin{eqnarray}
\zeta^{(i)}_j(s) = \epsilon_{ij}h_{i\rightarrow j}(\sigma_i(s-1))  \:.
\end{eqnarray}
The prefactor $\epsilon_{ij}$ determines whether the edge is symmetric or not: $\epsilon_{ij} = 1$ for undirected edges and $\epsilon_{ij}=0$ for directed edges.  We took a factorised initial distribution $P_0$: $P_0(\bsigma(0)) = \prod^N_{i=1}p_0(\sigma_i(0))$.
The single site marginal $P_{i, 0..t}$ is obtained  by summing $P_{0..t}$ in (\ref{eq:PathProb}) over all paths $\sigma^{0..t}_j$ with $j\neq i$ .  In general, we will use the notations 
\begin{eqnarray}
 \sigma_{S} = (\sigma_{i_1}, \sigma_{i_2}, \cdots, \sigma_{i_{|S|}}) \:,\\ 
P_S(\sigma^{0..t}_S|\btheta^{1..t}) = \sum_{\sigma^{0..t}_j,\: j\neq S}P(\bsigma^{0..t}|\btheta^{1..t})\:,
\end{eqnarray}
with $S$ a set of indices: $S = \left\{i_1, \cdots, i_{|S|}\right\}$, where $|S|$ denotes the size of the set $S$.  
Within this notation $P_{\partial_i, 0..t}$ is the joint probability of the paths on the neighbours of $i$.  When we sum  over all paths $\sigma^{0..t}_j$, with $j\neq i$, on the left hand and right hand side of (\ref{eq:PathProb}), we get
\begin{eqnarray}
 \fl P_{i, 0..t}\left(\sigma^{0..t}_i|\btheta\right) &= \sum_{\tau_{\partial_i}^{0..t}}P_{\partial_i\cup i, 0..t-1}\left(\tau_{\partial_i}^{0..t}, \tau^{0..t}_i\:\big|\:\btheta^{1..t}\right)
\nonumber \\ 
 \fl &= \sum_{\tau_{\partial_i}^{0..t}}P^{(i)}_{\partial_i, 0..t}\left(\tau_{\partial_i}^{0..t-1}\:\big|\: \btheta^{1..t-1}+ \bzeta^{(i), 1..t-1}\right) \prod^t_{s=1}W\left[\sigma_i(s)|h_i(s)\right] p_0(\sigma_i(0))  \label{eq:Pit}\:.
\end{eqnarray}
In the sequel we drop the subscript $i$ in the argument of  $P_{i, 0..t}$.
Now we make the Bethe-Peierls approximation: i.e. we assume that the spins in the neighbourhood $\partial_i$ of $i$ become independent when we remove the $i$-th spin: 
\begin{eqnarray}
\fl  P^{(i)}_{\partial_i, t}\left(\tau^{0..t}_{j_1}, \cdots, \tau^{0..t}_{j_{|\partial_i|}}| \btheta^{1..t}+ \bzeta^{(i), 1..t}\right) = \prod_{j\in\partial_i}P^{(i)}_{j, 0..t}\left(\tau^{0..t}_{j}|\theta^{1..t}_j + \epsilon_{ij}\zeta^{(i), 1..t}_{j}\right)\label{eq:BP} \:,
\end{eqnarray}
with $\partial_i = \left\{j_1, \cdots, j_{|\partial_i|}\right\}$.  In (\ref{eq:BP}) we took $\theta_j = 0$ when $j\notin \partial i \cup i$. We substitute (\ref{eq:BP}) in (\ref{eq:Pit}) to get the following set of recursive equations 
\begin{eqnarray}
 \fl P^{(\ell)}_{i, 0..t}\left(\sigma^{0..t}|\theta^{1..t}\right) = \sum_{\sigma^{0..t-1}_{\partial^{\rm in}_i\setminus \ell} }\:\left(\prod_{j\in\partial^{\rm{in}}_i\setminus \ell}P^{(i)}_{j, 0..t-1}\left(\sigma^{0..t-1}_{j}|\epsilon_{ij}\zeta^{(i), 1..t-1}_{j}\right)\right)
\nonumber \\ 
\left( \prod^t_{s=1}W\left[\sigma(s)|h^{(\ell)}_i(s)\right] p_{0}(\sigma(0))\right)\:,   \label{eq:DynCav}
\end{eqnarray}
 for the path probability $P^{(\ell)}_{i, 0..t}$ on the graph $G^{(\ell)}$, with $\ell \in \partial i$.  To derive (\ref{eq:DynCav}) we used $P^{(i,\ell)}_{j, 0..t}= P^{(i)}_{j, 0..t}$.   The set of $|E|$-equations (\ref{eq:DynCav}) determines the $|E|$-probability distributions  $P^{(\ell)}_{i, 0..t}\left(\sigma^{0..t}|\theta^{1..t}_i\right)$ at time step $t$ as a function of the  $|E|$-probability distributions  $P^{(\ell)}_{i, 0..t-1}(\sigma^{0..t-1}|\theta^{1..t-1}_i)$ at the previous time step $t-1$. In equation (\ref{eq:DynCav}) we only need to take the product over $j\in \partial^{\rm in}_{i}$ because the fields $h^{(\ell)}_i(s)$ depend only on $\sigma_{\partial^{\rm in}_i}$.
We call the equations (\ref{eq:DynCav}) the dynamical cavity equations analogous to the static equations (\ref{eq:CavFinal}).  The main difference is that (\ref{eq:DynCav}) are recursive equations of probabilities of paths propagating along the graph, while in (\ref{eq:CavFinal}) messages are propagated that determine the marginal probabilities of the stationary distribution.  Just like the static equations, the set of equations (\ref{eq:DynCav}) is exact on a tree. To find the marginal distributions $P_{i, 0..t}$  on the original graph $G$ from the cavity distributions, we need to combine equations (\ref{eq:Pit}) and (\ref{eq:BP}): 
\begin{eqnarray}
 \fl P_{i, 0..t}(\sigma^{0..t}|\theta^{1..t}) &= \sum_{\sigma^{0..t-1}_{\partial^{\rm in}_i}}\:\left(\prod_{j\in\partial^{\rm{in}}_i}P^{(i)}_{j, 0..t-1}\left(\sigma^{0..t-1}_{j}|\epsilon_{ij} \zeta^{(i), 1..t-1}_{j}\right) \right)
\nonumber \\ 
&\left(\prod^t_{s=1}W\left[\sigma(s)|h_i(s)\right] p_{0}(\sigma(0))\right)\:.  \label{eq:Marg}
\end{eqnarray}
Equations (\ref{eq:Marg}) are the dynamical versions of the set of equations (\ref{eq:CavMarg}). 
The initial problem of finding the single site marginals $P_{i, 0..t}$ from the $N$-site probability $P_{0..t}$ has a computational complexity $\mathcal{O}(2^N)$.  The set of equations (\ref{eq:DynCav}) and (\ref{eq:Marg}) has a linear complexity $\mathcal{O}(N)$ in the system size and an exponential complexity $\mathcal{O}\left(2^t\right)$ in time  which makes the dynamics solvable for a finite number of time steps.

The cavity equations simplify a lot when the graph is fully asymmetric.  In this case we can set $\epsilon_{ij} = 0$ in equation (\ref{eq:DynCav}). Therefore, the equations only have to be solved for $\theta^{0..t} = 0^{0..t}$, where $0^{0..t}$ is the null vector.  Moreover, because $\epsilon_{ij}=0$ the self-coupling disappears in (\ref{eq:DynCav}).  We can thus sum on the left and right hand side of (\ref{eq:DynCav}) over $(\sigma_i(0), \sigma_i(1), \cdots, \sigma_i(t-1))$ to get
\begin{eqnarray}
 \fl P^{(\ell)}_{i, t}(\sigma) =  \sum_{\sigma_j, \ j\in \partial^{\rm in}_i}\left(\prod_{j\in\partial^{\rm{in}}_i}P^{(i)}_{j, t-1}\left(\sigma_{j}\right)\right) W\left[\sigma|h\right]\:. \label{eq:DynCavAs}
\end{eqnarray}
Equation (\ref{eq:DynCavAs}) describes a  Markovian dynamics.
\section{The Ensemble Averaged Distribution of Paths}\label{sec:average}
We calculate the average of equation  (\ref{eq:DynCav}) over all links in the graph, i.e. all $a\in E$.  The graph is drawn from an ensemble of graphs $\mathcal{G}$.  We look at ensembles where the typical graphs have a local tree structure and the degrees on different sites are uncorrelated.   An example is the Poissonian ensemble $\mathcal{G}_p$ defined in (\ref{eq:EnsemblePoiss1}) of \ref{app:PoisDegreeI}.  The degree distribution is defined through a histogram as
\begin{eqnarray}
 \fl p_{\rm deg}(k^{\rm in}, k^{\rm out}, k^{\rm sym}) \equiv \frac{\sum^N_{i=1}\delta\left(k^{\rm{in}}; k^{\rm{in}}_i\right)\delta\left(k^{\rm{out}}; k^{\rm{out}}_i\right)\delta\left(k^{\rm sym}; k^{\rm sym}_i\right)}{N}\:. \label{eq:degreeDef}
\end{eqnarray}
In equation (\ref{eq:degreeDef}) we use the following notations: the indegree $k^{\rm{in}}_i = |\partial^{\rm{in}}_i|$, the outdegree $k^{\rm{out}}_i = |\partial^{\rm{out}}_i|$ and the symmetric degree $k^{\rm sym}_i = |\partial^{\rm{in}}_i \cap \partial^{\rm{out}}_i|$. For $N\rightarrow \infty$ the dynamics of Ising models on typical graphs drawn from such ensembles depends on the degree distribution (\ref{eq:degreeDef}). 
We define $\overline{P}^{\rm{d}}$ as the average of the path probabilities  $P^{(\ell)}_{i, t}$ over all directed edges $(i,\ell)$ of $E$:
\begin{eqnarray}
 \overline{P}^{\rm{d}}(\sigma^{0..t}) \equiv  \frac{\sum_{(i,\ell)\in E^{\rm{d}}}P^{(\ell)}_{i, 0..t}(\sigma^{0..t}|0^{1..t})}{|E^{\rm{d}}|} \:. \label{eq:defDir}
\end{eqnarray}
The average probability mass function $\overline{P}^{\rm{sym}}$ is defined as the average of $P^{(\ell)}_{i, t}$ over all links belonging to an undirected edge
\begin{eqnarray}
 \overline{P}^{\rm{sym}}\left(\sigma^{0..t}|\theta^{1..t}\right) \equiv  \frac{\sum_{(i, \ell) \in E^{\rm sym}}P^{(\ell)}_{i, 0..t}\left(\sigma^{0..t}|\theta^{1..t}\right)}{|E^{\rm sym}|} \:. \label{eq:defSym}
\end{eqnarray}
When we use the property that the spins in the neighbourhood of $i$ are uncorrelated, we can write
\begin{eqnarray}
  \overline{\prod_{j\in\partial_i\setminus \ell}P^{(i)}_{j, 0..t-1}\left(\sigma_{j}^{0..t-1}|\zeta^{(i), 1..t-1}_j\right)} =  \prod_{j\in\partial_i\setminus \ell}\overline{P}\left(\sigma_j^{0..t-1}| \zeta_{j}^{(i), 1..t-1}\right) \:. \label{eq:Sep}
\end{eqnarray} 
It is useful to focus on a specific example.  We consider fields of the type $h_{j\rightarrow i}(s-1) = J_{ji}\sigma_i(s-1)$, where the interactions strengths $J_{ij}$ are i.i.d.r.v. drawn from a distribution $R(J)$. When we take the average of the update equations (\ref{eq:DynCav}) according to the definitions (\ref{eq:defDir}) and (\ref{eq:defSym}), and use (\ref{eq:Sep}) we find the recursive equations for the averaged probability mass function of paths.   These recursive equations are given by: 
\begin{eqnarray}
 \fl \overline{P}^{\rm{d}}\left(\sigma^{0..t}\right) = \sum^{\infty}_{k^{\rm out}\geq 0}\sum^{\infty}_{k^{\rm in}\geq 0}\sum^{\rm{min}\it(k^{\rm out}, k^{\rm in})}_{k^{\rm sym}=0}\frac{p(k^{\rm in}, k^{\rm out},k^{\rm sym})(k^{\rm out}-k^{\rm sym})}{c_{\rm{out}}-c_{\rm {sym}}}
\nonumber \\ 
\fl\prod^{ k^{\rm in}}_{\ell = k^{\rm sym} + 1}\int dJ_{\ell}\: R(J_{\ell})\sum_{\sigma_{\ell}^{0..t-1}}\overline{P}^{\rm{d}}\left(\sigma_{\ell}^{0..t-1}\right)\prod^{ k^{\rm sym}}_{\ell=1} \int dJ_{\ell}\: R(J_{\ell})\sum_{\sigma_{\ell}^{0..t-1}}\overline{P}^{\rm{sym}}\left(\sigma_{\ell}^{0..t-1}|J_{\ell} \sigma^{0..t-2}\right)
\nonumber\\
p_0(\sigma(0))\prod^{t-1}_{s\geq 0}W\left[\sigma(s+1)|\theta(s) + \sum_{0<\ell'\leq k^{\rm in}}J_{\ell'}\sigma_{\ell'}(s)\right]\:, \label{eq:Pout}
\end{eqnarray}
and 
\begin{eqnarray}
 \fl \overline{P}^{\rm{sym}}\left(\sigma^{0..t}|\theta^{1..t}\right) &= \sum^{\infty}_{k^{\rm in}\geq 0}\sum^{k^{\rm in}}_{k^{\rm sym}=0}\frac{p(k^{\rm in}, k^{\rm sym}) k^{\rm sym}}{c_{\rm sym}}\prod^{ k^{\rm in}-1}_{\ell = k^{\rm sym}}\int dJ_{\ell}R(J_{\ell})\sum_{\sigma^{0..t-1}_{\ell}}\overline{P}^{\rm{d}}\left(\sigma_{\ell}^{0..t-1}\right)
\nonumber \\ 
&\prod^{ k^{\rm sym}-1}_{\ell=1} \int dJ_{\ell}R(J_{\ell})\sum_{\sigma_{\ell}^{0..t-1}}\overline{P}^{\rm{sym}}\left(\sigma_{\ell}^{0..t-1}|J_{\ell}\sigma^{0..t-2}\right)
\nonumber\\
&p_0(\sigma(0))\prod^{t-1}_{s\geq 0}W\left[ \sigma(s+1)|\theta(s+1) + \sum_{0<\ell'\leq k^{\rm in}-1}J_{\ell'}\sigma_{\ell'}(s)\right]\:.\label{eq:PSym}
\end{eqnarray}
We introduced the average connectivities $c_{\rm sym} \equiv \sum_{k^{\rm in}, k^{\rm out}, k^{\rm sym}}p(k^{\rm in}, k^{\rm out},k^{\rm sym})k^{\rm sym}$ and $c_{\rm out} \equiv \sum_{k^{\rm in}, k^{\rm out}, k^{\rm sym}}p(k^{\rm in}, k^{\rm out},k^{\rm sym})k^{\rm out}$.

The averaged probability mass function $\overline{P}^{\rm real}(\sigma^{0..t})$ over the marginals $P_i(\sigma^{0..t})$, defined through $\overline{P}^{\rm real}(\sigma^{0..t}) \equiv \sum_{i}P_i(\sigma^{0..t})/N$, can be calculated from (\ref{eq:Marg}):
\begin{eqnarray}
 \fl \overline{P}^{\rm{real}}\left(\sigma^{0..t}|\theta^{1..t}\right)  &= \sum^{\infty}_{k^{\rm in}\geq 0}\sum^{k^{\rm in}}_{k^{\rm sym}=0}p(k^{\rm in}, k^{\rm sym})  \prod^{ k^{\rm in}}_{\ell = k^{\rm sym}+1}\int dJ_{\ell}R(J_{\ell})\sum_{\sigma^{0..t-1}_{\ell}}\overline{P}^{\rm{d}}\left(\sigma_{\ell}^{0..t-1}\right)   
\nonumber \\ 
\fl &\prod^{k^{\rm sym}}_{\ell= 1} \int dJ_{\ell}R(J_{\ell})\sum_{\sigma^{0..t-1}_{\ell}}\overline{P}^{\rm{sym}}\left(\sigma_{\ell}^{0..t-1}|J_{\ell}\sigma^{0..t-2}\right)
\nonumber\\
\fl  &
p_0(\sigma(0))\prod^{t-1}_{s\geq 0}W\left[\sigma(s+1)|\theta(s+1) + \sum_{0<\ell'\leq k^{\rm in}}J_{\ell'}\sigma_{\ell'}(s)\right] \:.\label{eq:PReal}
\end{eqnarray}
The Markovian dynamics of $N$ spins defined in (\ref{eq:W})  is thus reduced to an effective non-Markovian dynamics of one single spin given by the recursive equations (\ref{eq:Pout}), (\ref{eq:PSym}) and (\ref{eq:PReal}).   Equations analogous to (\ref{eq:Pout}) and (\ref{eq:PSym}) were derived in \cite{Mimura2} in the context of LDGM channel coding using the generating functional analysis.

For fully asymmetric graphs, see (\ref{eq:DynCavAs}), we remark that $P^{(\ell)}_{i, t}(\sigma) = P_{i, t}(\sigma)$, but the averages,  $\overline{P}^{\rm{d}}_t\left(\sigma\right)  \equiv \overline{P^{(\ell)}_{i, t}(\sigma)}$ and $ \overline{P}^{\rm{real}}_t\left(\sigma\right)  \equiv \overline{P_{i, t}(\sigma)}$, over, respectively, the links and the sites are different.  Indeed:
\begin{eqnarray}
  \fl  \overline{P}^{\rm{d}}_t\left(\sigma\right) =\sum_{k^{\rm in}}p(k^{\rm in})\frac{c(k^{\rm in})}{c_{\rm{out}}} \prod_{0<\ell\leq k^{\rm in}} \int dJ_{\ell}R(J_{\ell})\sum_{\sigma_{\ell}}\overline{P}^{\rm{d}}_{t-1}(\sigma_{\ell})  \nonumber 
 \\ 
p(\sigma(0))\prod^t_{s=1}W\left[\sigma(s)|h(s)\right]\:,  \label{eq:POutAs}\\ 
 \fl \overline{P}^{\rm{real}}_t\left(\sigma\right) =\sum_{k^{\rm in}}p(k^{\rm in})\prod_{0<\ell\leq k^{\rm in}}\int dJ_{\ell}R(J_{\ell})\sum_{\sigma_{\ell}}\overline{P}^{\rm{d}}_{t-1}(\sigma_{\ell})\:p(\sigma(0))\prod^t_{s=1}W\left[\sigma(s)|h(s)\right] \:,  \label{eq:PRealAs}
\end{eqnarray}
with $c(k^{\rm in}) = \sum_{k^{\rm out}}p(k^{\rm out}|k^{\rm in})k^{\rm out}$ and $c_{\rm out} = \sum_{k^{\rm out}}p(k^{\rm out})k^{\rm out}$.

\section{Examples of Dynamics}\label{sec:bench}
In this section we define the type of dynamics we study by specifying the form of the transition probabilities $W\left[\sigma|h\right]$ used in equation (\ref{eq:W}). 
\subsection{Glauber Dynamics} \label{subsec:Glauber}
We consider Glauber dynamics for an Ising model with $n=1$, i.e. $\sigma_i\in \left\{-1,1\right\}$.  Every spin $\sigma_i(t)$ evolves under the influence of the field $h_i(t-1)$ with a transition probability $W_{\rm{g}}(\sigma_i(t)|h_i(t))$ defined through:
\begin{eqnarray}
 W_{\rm{g}}[\sigma|h] \equiv \frac{\exp\left(\beta \sigma h\right)}{2\cosh\left(\beta h\right)} \label{eq:GlauberSigma}\:.
\end{eqnarray}
The parameter $\beta$ is the inverse of the temperature $T$.
It is possible to implement the dynamics defined by (\ref{eq:GlauberSigma}) and  (\ref{eq:W}) with the heat-bath algorithm \cite{Binder2005}.  When the graph is fully symmetric detailed balance is satisfied and the Hamiltonian is given by equation (\ref{eq:HamEquil}).

As an example we apply the recursive equations (\ref{eq:Pout}), (\ref{eq:PSym}) and (\ref{eq:PReal}) to a Poissonian ensemble $\mathcal{G}_p$ defined by the probability $P_{\rm p}(\bC; c, \epsilon)$ of the connectivity matrix $\bC$:  
\begin{eqnarray}
\fl  P_{\rm p}(\bC; c, \epsilon) = \prod_{i<j}\left(\frac{c}{N}\delta(c_{ij}; 1) + \left(1-\frac{c}{N}\right)\delta(c_{ij}; 0) \right) 
\nonumber \\  \prod_{i>j}\left(\epsilon \delta(c_{ij}; c_{ji}) + \left(1-\epsilon\right)\left(\frac{c}{N}\delta(c_{ij}; 1) + \left(1-\frac{c}{N}\right)\delta(c_{ij}; 0) \right) \right) \:. \label{eq:DefPoiss}
\end{eqnarray}
The parameter $c$ controls the number of edges while $\epsilon$ controls the fraction of symmetric edges in the graph.
The $\delta(a;b)$ are Kronecker delta functions.  
For $N\rightarrow \infty$, the typical degree distribution of a graph drawn from this ensemble, $p_{\rm p}(k^{\rm in}, k^{\rm out},  k^{\rm{sym}})$, is given by:
\begin{eqnarray}
 \fl p_{\rm p}(k^{\rm in}, k^{\rm out}, k^{\rm sym}) &=\left(\exp\left[-c\epsilon\right] \frac{\left(c\epsilon\right)^{k^{\rm sym}}}{k^{\rm sym}!} \right) \left(\exp\left[-\left(1-\epsilon\right)c\right]\frac{\left(c\left(1-\epsilon\right)\right)^{k^{\rm in}-k^{\rm sym}}}{(k^{\rm in}-k^{\rm sym})!}\right)
\nonumber \\  
&\left(\exp\left[-\left(1-\epsilon\right)c\right] \frac{\left(c\left(1-\epsilon\right)\right)^{k^{\rm out}-k^{\rm sym}}}{(k^{\rm out}-k^{\rm sym})!}\right) \:. \label{eq:PoisDegree}
\end{eqnarray}
For the derivation of (\ref{eq:PoisDegree}) we refer to \ref{app:PoisDegreeI}.
Substitution of (\ref{eq:PoisDegree}) in (\ref{eq:PSym}) and (\ref{eq:PReal}) leads to,
\begin{eqnarray}
 \fl \overline{P}^{\rm{real}}(\sigma^{0..t}|\theta^{1..t}) &= \overline{P}^{\rm{sym}}(\sigma^{0..t}|\theta^{1..t})  = \sum_{k \geq 0}\frac{e^{-c}c^k}{k!} \int \left(\prod^{k}_{\ell=1} dJ_\ell R(J_{\ell})\right) \nonumber \\ 
\fl&  \prod^k_{\ell=1} \sum_{\sigma^{0..t-1}_\ell}\left[\epsilon \overline{P}^{\rm sym}\left(\sigma^{0..t-1}_\ell|J_\ell\sigma^{0..t-2}\right) +(1-\epsilon)\overline{P}^{\rm sym}(\sigma^{0..t-1}_\ell|0^{0..t-1})\right] 
\nonumber \\ 
\fl&  p(\sigma(0)) \prod^{t-1}_{s\geq 0} \frac{\exp\left[\beta \sigma(s+1)\left(\theta(s+1) + \sum_{0<\ell' \leq k}J_{\ell'} \sigma_{\ell'}(s)\right)\right]}{2\cosh\left[\beta \left(\theta(s+1) + \sum_{0<\ell' \leq k}J_{\ell'} \sigma_{\ell'}(s)\right)\right]} \:. \label{eq:PoissGraph}
\end{eqnarray}
For the Poissonian ensemble equation (\ref{eq:change}) is valid such that $\overline{P}^{d}(\sigma^{0..t}) = \overline{P}^{\rm sym}(\sigma^{0..t}|0^{1..t-1}) = \overline{P}^{\rm real}(\sigma^{0..t})$.
We find equation (\ref{eq:PoissGraph}) which is identical to the main result of reference \cite{Hat2004} derived by calculating the generating function.  Hence, we conclude that the equations (\ref{eq:Pout}), (\ref{eq:PSym}) and (\ref{eq:PReal}) are consistent with the results found in \cite{Hat2004}.

\subsection{Coupled Dynamics}\label{subsec:coupled}
We define a dynamics of two sets of $N$ spins under the influence of the same thermal noise.  We thus have $n=2$, i.e. the dynamic variables are $(\sigma_i(t), \tau_i(t)) \in \left\{-1, 1\right\}\times\left\{-1, 1\right\}$.  The spins $(\sigma_i(t), \tau_i(t))$ feel only the influence of their neighbouring spins $(\sigma_i, \tau_i)$, with $i\in\partial^{\rm in}_i$, through, respectively, the fields $h_{i}(t)$ and $g_{i}(t)$. The fields $h_{i}(t)$ and $g_i(t)$ depend, respectively, only on the sets $\bsigma(t-1)$ and $\btau(t-1)$.  The spins $(\sigma_i(t), \tau_i(t))$ evolve according to $W_{\rm{c}}\left[(\sigma_i(t),\tau_i(t)); h_i(t),g_i(t)\right]$ :
\begin{eqnarray}
 \fl W_{\rm{c}}\left[(\sigma,\tau)| h,g\right] \equiv\delta\left(\sigma;-\tau\right)|r_h-r_g| 
\nonumber \\ 
\fl + \delta\left(\sigma;\tau\right)(1-|r_h-r_g|)\Theta(r_h-r_g)\left[\delta(\sigma;1)\frac{r_g}{1 + r_g -r_h} + \delta(\sigma; -1)\frac{1-r_h}{1 + r_g -r_h}\right]
\nonumber \\ 
\fl + \delta\left(\sigma;\tau\right)(1-|r_h-r_g|)\Theta(r_g-r_h)\left[\delta(\sigma;1)\frac{r_h}{1 + r_h -r_g} + \delta(\sigma; -1)\frac{1-r_g}{1 + r_h -r_g}\right]\:, \label{eq:defWC}
\end{eqnarray}
where $\Theta$ is the Heaviside step function and the weights $r_h$ and $r_g$ are given by
\begin{eqnarray}
r_h = \frac{\exp\left(\beta h\right)}{2\cosh\left(\beta h\right)}  \:, \ \ 
r_g = \frac{\exp\left(\beta  g\right)}{2\cosh\left(\beta g\right)}  \:. 
\end{eqnarray}
Equation (\ref{eq:defWC}) can be simulated using a heat-bath algorithm where at each time step we choose the same random numbers for both set of spins $\bsigma$ and $\btau$.  A more compact form of $W_{\rm{c}}$ is: 
\begin{eqnarray}
\fl W_{\rm{c}}\left[(\sigma,\tau)| h,g\right] &=\delta\left(\sigma;-\tau\right)|r_h-r_g| 
\nonumber \\
&+ \delta\left(\sigma;\tau\right)\Theta(r_h-r_g)\left[\delta(\sigma;1)\:r_g + \delta(\sigma; -1)\left(1-r_h\right)\right]
\nonumber \\ 
&+ \delta\left(\sigma;\tau\right)\Theta(r_g-r_h)\left[\delta(\sigma;1)\:r_h + \delta(\sigma; -1)\left(1-r_g\right)\right] \label{eq:WDynErgo}
\:.
\end{eqnarray}

 When the thermal average of the distance between the paths $\bsigma(t)$ and $\btau(t)$ does not converge to zero for $t\rightarrow \infty$, even when the initial distance between $\bsigma(0)$ and $\btau(0)$ is very small, the system is in a chaotic phase. We use the transition probability $W_{\rm{c}}$  to determine the phase transitions to this chaotic phase.  Chaotic behaviour has been studied in  \cite{Der1987} for spin glasses and in \cite{Der1987Co} for neural networks.  The coupled dynamics (\ref{eq:WDynErgo}) can not satisfy detailed balance.  

\section{The Path Entropy and the Distribution of the Probability Distributions of Paths}\label{sec:ensemble}
The fluctuations of the path probabilities $P^{(\ell)}_{i, 0..t}$ over all links are given by the distribution of the probabilities of the paths which we will call $\mathcal{P}$.   They determine quantities like the average path entropy $\overline{\mathcal{S}}(t)$.  On the basis of the recursive equations for the distributions $\mathcal{P}$ we discuss in section \ref{sec:9} the stationary solutions of the dynamics.  

The average path entropy is defined as
\begin{eqnarray}
\overline{\mathcal{S}}(t) \equiv -\overline{\sum_{\bsigma^{0..t}}P_{0..t}(\bsigma^{0..t})\log\left(P_{0..t}(\bsigma^{0..t})\right) } \:, \label{eq:defPath}
\end{eqnarray}
where the bar denotes the average over the quenched variables.  
With the cavity method \cite{Mez2001}, we can write 
\begin{eqnarray}
 \overline{\mathcal{S}}(t) = \sum^N_{i=1}\Delta \mathcal{S}^{\rm site}_{i}(t)- \frac{1}{2} \sum_{a \in E^{\rm sym} }\: \Delta \mathcal{S}^{\rm link}_{a}(t) - \sum_{a \in E^{\rm{d}} }\: \Delta \mathcal{S}^{\rm link}_{a}(t) \:. \label{eq:pathEntropy2}
\end{eqnarray}
The quantity $\Delta \mathcal{S}^{\rm site}_{i}(t) $ is the increment in the entropy $\mathcal{S}$ when the $i$-th site is added to the graph $G^{(i)}$: 
\begin{eqnarray}
 \fl \Delta \mathcal{S}^{\rm site}_{i}(t) = 
- \sum_{\sigma^{0..t}}\ \sum_{\sigma^{0..t}_{\partial_i}}P_{\partial_i, 0..t}(\sigma^{0..t}_{\partial_i})\prod^t_{s=1}W\left[\sigma(s)|h(s)\right]p_0(\sigma(0)) 
\nonumber \\ 
\log\left( P_{\partial i, 0..t}(\sigma^{0..t}_{\partial_i })\prod^t_{s=1}W\left[\sigma(s)|h(s)\right]p_0(\sigma(0)) \right) \:, \label{eq:siteEntropy}
\end{eqnarray}
with
\begin{eqnarray}
\fl P_{\partial_i, 0..t }(\sigma^{0..t}_{\partial_i}) = \prod_{(j,i)\in E^{\rm{d}}  }P^{(i)}_{j, 0..t}(\sigma^{0..t}_j)\prod_{(j, i)\in E^{\rm sym}}P^{(i)}_{j, 0..t}(\sigma^{0..t}_j|J_{ij}\sigma^{0..t-1})\:.
\end{eqnarray}
The quantity $\Delta \mathcal{S}^{\rm link}_{a}(t) $ is minus the entropy difference when remove the link $a$ from the graph $G$ 
\begin{eqnarray}
\fl  \Delta \mathcal{S}^{\rm link}_{(i,j)}(t) = -  \sum_{\sigma^{0..t},\: \tau^{0..t}} P_{i, 0..t}(\sigma^{0..t}) P^{(i)}_{j, 0..t}(\tau^{0..t}|J_{ij}\sigma^{0..t-1})
\nonumber \\
\log\left(P_{i, 0..t}(\sigma^{0..t}) P^{(i)}_{j, 0..t}(\tau^{0..t}|J_{ij}\sigma^{0..t-1})\right)\:.\label{eq:linkEntropy}
\end{eqnarray}
The summation over the sites in equation (\ref{eq:pathEntropy2}) can be done when we know the distributions of the probabilities of paths $\mathcal{P}$ on the graph.

We define the following distributions
\begin{eqnarray}
\fl  \mathcal{P}^{\rm{d}}\left(P\right) &\equiv \frac{\sum_{(i, \ell)\in E^{\rm{d}}} \prod_{\sigma^{0..t}}\delta\left(P(\sigma^{0..t}) - P^{(\ell)}_{i, 0..t}(\sigma^{0..t}) \right)}{|E^{\rm{d}}|} \:, \\ 
 \fl \mathcal{P}^{\rm{sym}}\left(P\right) &\equiv \frac{\sum_{(i, \ell)\in E^{\rm sym}} \prod_{\sigma^{0..t},\theta^{1..t}}\delta\left(P(\sigma^{0..t}|\theta^{1..t}) - P^{(\ell)}_{i, 0..t}(\sigma^{0..t}|\theta^{1..t}) \right)}{|E^{\rm sym}|} \:, \\ 
\fl \mathcal{P}^{\rm{real}}\left(P\right) &\equiv \frac{\sum^N_{i=1}\prod_{\sigma^{0..t}}\delta\left(P(\sigma^{0..t}) - P_i(\sigma^{0..t}) \right)}{N} \:.
\end{eqnarray}
When the variables $J_{ij}$ are i.d.d.r.v. $\mathcal{P}^{\rm d}$ satisfies the recursive equation: 
\begin{eqnarray}
 \fl \mathcal{P}^{\rm{d}}\left(P\right) &= \sum^{\infty}_{k^{\rm out}\geq 0}\sum^{\infty}_{k^{\rm in}\geq 0}\sum^{\rm{min}\it(k^{\rm out}, k^{\rm in})}_{k^{\rm sym}=0}\frac{p(k^{\rm in}, k^{\rm out},k^{\rm sym})(k^{\rm out}-k^{\rm sym})}{c_{\rm{out}}-c_{\rm sym}}
\nonumber \\ 
\fl 
&\prod^{k^{\rm sym}}_{\ell=1} \int dJ_{\ell}R(J_{\ell})\int dP_\ell \: \mathcal{P}^{\rm{sym}}\left(P_\ell\right)\prod^{ k^{\rm in}}_{\ell = k^{\rm sym} +1}\int dJ_{\ell}R(J_{\ell})\int dP_\ell \: \mathcal{P}^{\rm{out}}\left(P_\ell\right) 
\nonumber\\
 \fl&\prod_{\sigma^{0..t}}\delta\Bigg(P(\sigma^{0..t})- \mathcal{F}^{\rm{d}}\left(\sigma^{0..t}; \left\{P_{\ell'}, J_{\ell'}\right\}_{\ell'=1..k^{\rm in}}\right)\Bigg)\:, \label{eq:distriout}
\end{eqnarray}
with 
\begin{eqnarray}
\fl  \mathcal{F}^{\rm{d}}\left(\sigma^{0..t}; \left\{P_\ell, J_\ell\right\}_{\ell=1..k^{\rm in}}\right) = \sum_{\sigma^{0..t-1}_1,\cdots,\sigma^{0..t-1}_{k^{\rm in}}}\prod^{k^{\rm sym}}_{\ell = 1}P_\ell(\sigma^{0..t-1}_\ell|J_\ell \sigma^{0..t-2})
\nonumber \\ 
\fl  \prod^{k^{\rm in}}_{\ell = k^{\rm sym} +1 }P_\ell(\sigma^{0..t-1}_\ell) \: \left(p_0(\sigma(0))\prod^{t-1}_{s\geq 0}\frac{\exp\left[\beta \sigma(s+1) \sum_{0<\ell'\leq k^{\rm in}}J_{\ell'}\sigma_{\ell'}(s)\right]}{2\cosh\left[\beta \sum_{0<\ell'\leq k^{\rm in}}J_{\ell'}\sigma_{\ell'}(s)\right]}\right) \:.\label{eq:FD}
\end{eqnarray}
The distribution along symmetric links is given by $\mathcal{P}^{\rm sym}$
\begin{eqnarray}
 \fl \mathcal{P}^{\rm{sym}}\left(P\right) &=  \sum^{\infty}_{k^{\rm in}\geq 0}\sum^{k^{\rm in}}_{k^{\rm sym}=0}\frac{p(k^{\rm in}, k^{\rm sym})k^{\rm sym}}{c_{\rm sym}}
\nonumber \\ 
\fl 
&\prod^{ k^{\rm sym}-1}_{\ell=1} \int dJ_{\ell}R(J_{\ell})\int dP_\ell\mathcal{P}^{\rm{sym}}\left(P_\ell\right)\prod^{ k^{\rm in}-1}_{\ell = k^{\rm sym}}\int dJ_{\ell}R(J_{\ell})\int dP_\ell\mathcal{P}^{\rm{out}}\left(P_\ell\right) 
\nonumber\\
 \fl &\prod_{\sigma^{0..t},\theta^{1..t}}\delta\Bigg(P\left(\sigma^{0..t}|\theta^{1..t}\right)-\mathcal{F}^{\rm sym}\left(\sigma^{0..t}|\theta^{1..t}; \left\{P_{\ell'}, J_{\ell'}\right\}_{\ell'=1..k^{\rm in}-1}\right)\Bigg)\:,\label{eq:distrisym}
\end{eqnarray}
with 
\begin{eqnarray}
\fl \mathcal{F}^{\rm sym}\left(\sigma^{0..t}|\theta^{1..t}; \left\{P_\ell, J_\ell\right\}_{\ell=1..k^{\rm in}-1}\right) = \sum_{\sigma^{0..t-1}_1,\cdots,\sigma^{0..t-1}_{k^{\rm in}-1}}\prod^{k^{\rm sym}-1}_{ \ell=1}P_\ell(\sigma^{0..t-1}_\ell|J_\ell \sigma^{0..t-2}_\ell)
\nonumber \\ 
\fl \prod^{k^{\rm in}-1}_{\ell = k^{\rm sym}}P_\ell(\sigma^{0..t-1}_\ell) \left(p_0(\sigma(0))\prod^{t-1}_{s\geq 0}\frac{\exp\left[\beta \sigma(s+1)\left(\theta(s+1) + \sum_{0<\ell'\leq k^{\rm in}-1}J_{\ell'}\sigma_{\ell'}(s)\right)\right]}{2\cosh\left[\beta\theta(s+1) +\beta\sum_{0<\ell'\leq k^{\rm in}-1}J_{\ell'}\sigma_{\ell'}(s)\right]}\right)\:.\nonumber \\  \label{eq:FSym}
\end{eqnarray}
The distribution of the single site marginals $P_i$ on the original graph is given by
\begin{eqnarray}
 \fl \mathcal{P}\left(P\right) = \sum^{\infty}_{k^{\rm in}\geq 0}\sum^{k^{\rm in}}_{k^{\rm sym}=0}p(k^{\rm in}, k^{\rm sym})\prod^{k^{\rm sym}}_{\ell=1} \int dJ_{\ell}R(J_{\ell})\int dP_\ell\mathcal{P}^{\rm{sym}}\left(P_\ell\right)
\nonumber \\ 
\fl 
\prod^{k^{\rm in}}_{\ell = k^{\rm sym}+1}\int dJ_{\ell}R(J_{\ell})\int dP_\ell\mathcal{P}^{\rm{d}}\left(P_\ell\right)  \prod_{\sigma^{0..t}}\delta\Bigg(P(\sigma^{0..t})-\mathcal{F}^{\rm real}\left(\sigma^{0..t} ;\left\{P_{\ell'}, J_{\ell'}\right\}_{\ell'=1..k^{\rm in}}\right) \Bigg)
\label{eq:distrireal}
\end{eqnarray}
with 
\begin{eqnarray}
\fl \mathcal{F}^{\rm real}\left(\sigma^{0..t}; \left\{P_\ell, J_\ell\right\}_{\ell=1..k^{\rm in}}\right) =  \sum_{\sigma^{0..t-1}_1,\cdots,\sigma^{0..t-1}_{k^{\rm in}}}\prod^{k^{\rm sym}}_{\ell=1}P_\ell(\sigma^{0..t-1}_\ell|J_\ell \sigma^{0..t-2}_\ell)
\nonumber \\ 
\fl \prod^{k^{\rm in}}_{\ell = k^{\rm sym}+1}P_\ell(\sigma^{0..t-1}_\ell)\left(p_0(\sigma(0))\prod^{t-1}_{s\geq 0}\frac{\exp\left[\beta \sigma(s+1) \sum_{0<\ell'\leq k^{\rm in}}J_{\ell'}\sigma_{\ell'}(s)\right]}{2\cosh\left[ \beta\sum_{0<\ell'\leq k^{\rm in}}J_{\ell'}\sigma_{\ell'}(s)\right]}\right)\:.\label{eq:FReal}
\end{eqnarray}
In section \ref{sec:9} we use the equations (\ref{eq:distriout}), (\ref{eq:distrisym}) and (\ref{eq:distrireal}) to derive the stationary limit from the dynamics. When we compare the equations (\ref{eq:distriout}) and (\ref{eq:distrisym}) with the density evolution equations (\ref{eq:WU}) we see a couple of differences.  Since the graph is directed we have now two distributions: one for the probabilities propagating along symmetric edges and one for the probabilities propagating along directed edges.  Since the equations (\ref{eq:distriout}) and (\ref{eq:distrisym}) describe the dynamics of the model they are recursive equations.   The computational complexity of (\ref{eq:distriout}) and (\ref{eq:distrisym}) scales exponentially in time.  Equation (\ref{eq:distrireal}) is the dynamical equivalent of (\ref{eq:WR}).

\begin{figure}[h!]
\begin{center}
\includegraphics[angle=-90, width=.6 \textwidth]{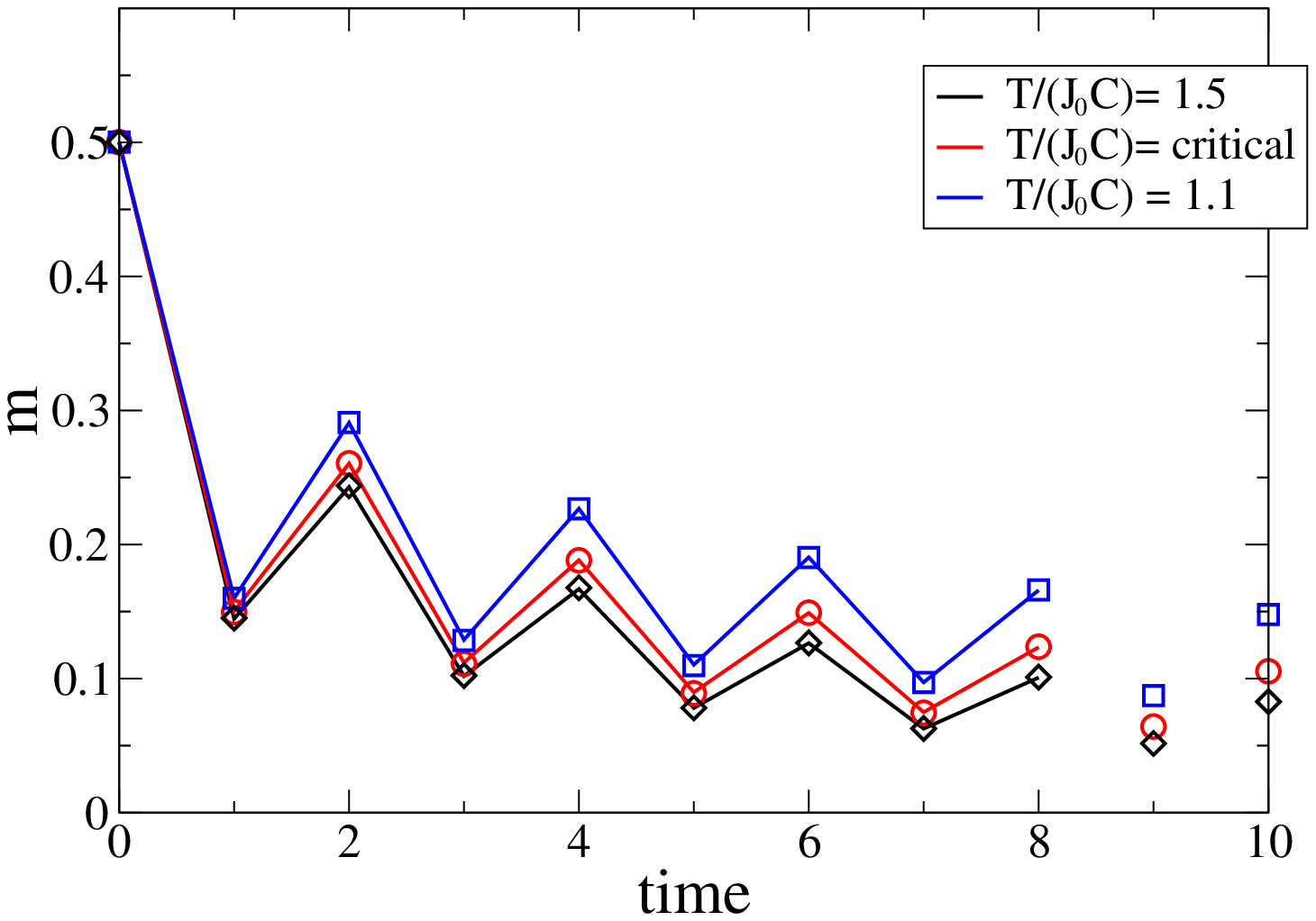}
\caption{The magnetisation $m$ as a function of discrete time for an Ising model on a symmetric Bethe lattice with connectivity $C=3$.  The interactions are drawn from the bimodal distribution (\ref{eq:distriRho}) with $\rho =0.25$.  The exact enumeration of the recursive equations (\ref{eq:PSym}) and (\ref{eq:PReal}) (lines) are compared with Monte Carlo simulations (markers).  The red line is calculated at the critical temperature $T_{\rm{c}} \approx 1.3459$. \label{fig:BetheFirstTime}}
\end{center}
\begin{center}
\includegraphics[angle=-90, width=.6 \textwidth]{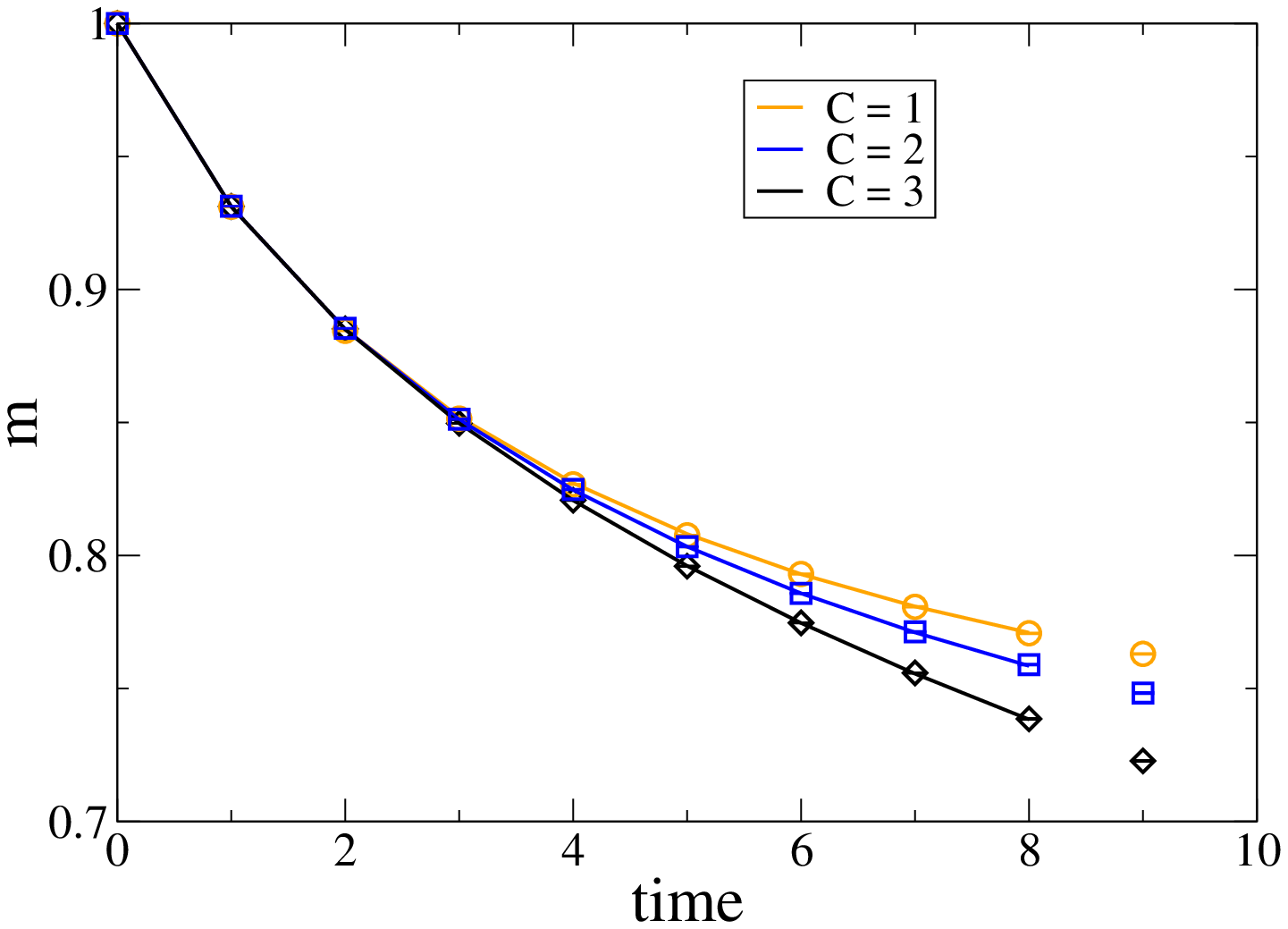}
\caption{The time evolution of the magnetisation $m$ at $T=1.8$ for an Ising model on a Bethe lattice without bond disorder for different levels of asymmetry.    Results are shown for graphs with fixed indegree $k^{\rm in}= 3$ and a given fixed outdegree $k^{\rm out} = C$.   The exact enumeration of the recursive equations (\ref{eq:Pout}), (\ref{eq:PSym}) and (\ref{eq:PReal}) (lines) are compared with Monte Carlo simulations (markers). } \label{fig:BPC3}
\end{center}
\end{figure}

\section{Comparison with Simulations}\label{sec:simulation}
In this section we compare the magnetisation $m(t) = \sum_{\sigma^{0..t}}\sigma(t)\overline{P}^{\rm{real}}\left(\sigma^{0..t}|0^{1..t}\right)$, predicted by equations (\ref{eq:Pout}), (\ref{eq:PSym}) and (\ref{eq:PReal}), with results from simulations.  It is difficult to develop an Eisfeller-Opper scheme \cite{Eiss1992} for these equations because the probability distributions of paths $\overline{P}^{\rm sym}$ depend on the fields $\theta^{1..t}$, such that it is necessary to solve (\ref{eq:PSym}) for all $2^t$ possible values of $\theta^{1..t}$.  We calculate the first time steps through an exact enumeration of the equations (\ref{eq:Pout}), (\ref{eq:PSym}) and (\ref{eq:PReal}).  

In figures \ref{fig:BetheFirstTime}  and \ref{fig:BPC3} we compare the magnetisation in the first time steps with the results obtained through Monte Carlo simulations.  In figure \ref{fig:BetheFirstTime} we present the results for an Ising model with bond disorder on a symmetric Bethe lattice. The interactions $J$ are drawn from the bimodal distribution~$R$:
\begin{eqnarray}
R(J) = \left(\frac{1+\rho}{2}\right)\delta(J-J_0) +   \left(\frac{1-\rho}{2}\right)\delta(J+J_0)\:,  \label{eq:distriRho}
\end{eqnarray}
with $\rho$ the bias in the couplings.
 In figure \ref{fig:BPC3} we show results for the Ising model on a Bethe lattice without bond disorder.  Both, the exact enumeration and the simulation give consistent results confirming the correctness of equations  (\ref{eq:Pout}), (\ref{eq:PSym}) and (\ref{eq:PReal}).


\section{The Ising Model on a Fully Asymmetric Bethe Lattice}\label{sec:Bethe}
The dynamical cavity equations (\ref{eq:DynCav}) simplify to (\ref{eq:DynCavAs}) for fully asymmetric graphs.  We illustrate this for the Ising model on a graph with a degree distribution $p_B(k^{\rm in}, k^{\rm out}, k^{\rm sym}) = \delta(k^{\rm in}-C)p(k^{\rm out})\delta(k^{\rm sym})$.  We call the graphs drawn from this ensemble asymmetric Bethe lattices.   Because now $\overline{P}^{\rm real}_t(\sigma) = \overline{P}^{\rm d}_t(\sigma)$, we only need to solve the recursive equation (\ref{eq:POutAs}).  We discuss this model for two typical dynamics.
\subsection{Glauber Dynamics}
In this subsection we let the spins evolve through Glauber dynamics defined in section \ref{subsec:Glauber}.  The probability of $\sigma_i(s)$ is given by equation (\ref{eq:GlauberSigma}) with a field $h_i(s) = \sum_{j\in \partial^{\rm in}_i}J_{ji}\sigma_j(s-1)$ .
\subsubsection{Iterative Equations}$\\$ 
We derive the equations that give the evolution over time of the following macroscopic observables: the average magnetisation, the correlation function and the distribution of magnetisations.

We define the magnetisation $m(t)$ through the relation $P_t(\sigma) = (m(t)\sigma+1)/2$.  From equation (\ref{eq:POutAs}) we  then get for the evolution of the magnetisation under Glauber dynamics
\begin{eqnarray}
 \fl m(t+1) = \prod_{0<\ell\leq C}\left[\sum_{\sigma_{\ell}}\left(\frac{1+\sigma_{\ell}m(t)}{2}\right)\right]
\Bigg\langle \tanh\left[\beta\left(\theta(t) + \sum_{0<\ell\leq C}J_{\ell}\sigma_\ell\right)\right]\Bigg\rangle_{J_1,J_2,\cdots,J_{C}}\:. \label{eq:mAs}
\end{eqnarray}

To find the correlation function $C(t, t')$ between spins at time step $t$ and $t'$, we sum equation (\ref{eq:Pout}) over all spins except $\sigma(t)$ and $\sigma(t')$.   We get the recursive equation for the two time marginal $P_{t,t'}(\sigma(t), \sigma(t'))$.  When we define $C(t, t')$ through
\begin{eqnarray}
 P_{t,t'}(\sigma, \tau) = \frac{1}{4}\left[1 + m(t)\sigma + \tau m(t') + C(t,t')\sigma \tau\right] \:,
\end{eqnarray}
we obtain the recursive equation for the correlation function: 
\begin{eqnarray}
  \fl C(t+1,t'+1) =     \prod_{0<\ell\leq C}\left(\sum_{\sigma_\ell\tau_\ell}\frac{\left[1+\sigma_\ell m(t) + \tau_\ell m(t') + \sigma_\ell\tau_\ell C(t,t'\right]}{4}\right) 
\nonumber \\ 
\fl  \Bigg\langle \tanh\left[\beta\left(\theta(t) + \sum_{0<\ell\leq C}J_{\ell}\sigma_\ell\right)\right]\tanh\left[\beta\left(\theta(t') + \sum_{0<\ell\leq C}J_{\ell}\tau_\ell\right)\right]\Bigg\rangle_{J_1,J_2,\cdots,J_{C}} \label{eq:corr}\:.
\end{eqnarray}

When we calculate the distribution of the marginals $P^{(\ell)}_{i, t}$ from equation (\ref{eq:DynCavAs}), or equivalently the distribution $W_t(m)$ of the corresponding magnetisations, the following recursive equation appears
\begin{eqnarray}
\fl  W_t(m) = \int \prod^C_{\ell=1}\left(dJ_\ell R(J_{\ell})\right) \int \prod^C_{\ell=1}\left(dm_\ell W_{t-1}(m_\ell)\right)
\nonumber \\ 
 \delta\left( m -
\prod_{0<\ell\leq C}\sum_{\sigma_{\ell}}\left(\frac{1+\sigma_{\ell}m_\ell}{2}\right)
\tanh\left[\beta\left(\theta(t) + \sum_{0<\ell'\leq C}J_{\ell'}\sigma_{\ell'}\right)\right] \right)\:. \label{eq:Wm}
\end{eqnarray}
The time evolution determined by the equations (\ref{eq:mAs}), (\ref{eq:corr}) and (\ref{eq:Wm} is confirmed  by numerical simulations.  
\subsubsection{A Stationary Solution}
\begin{figure}[h!]
\begin{center}
\includegraphics[angle=-90, width=.6 \textwidth]{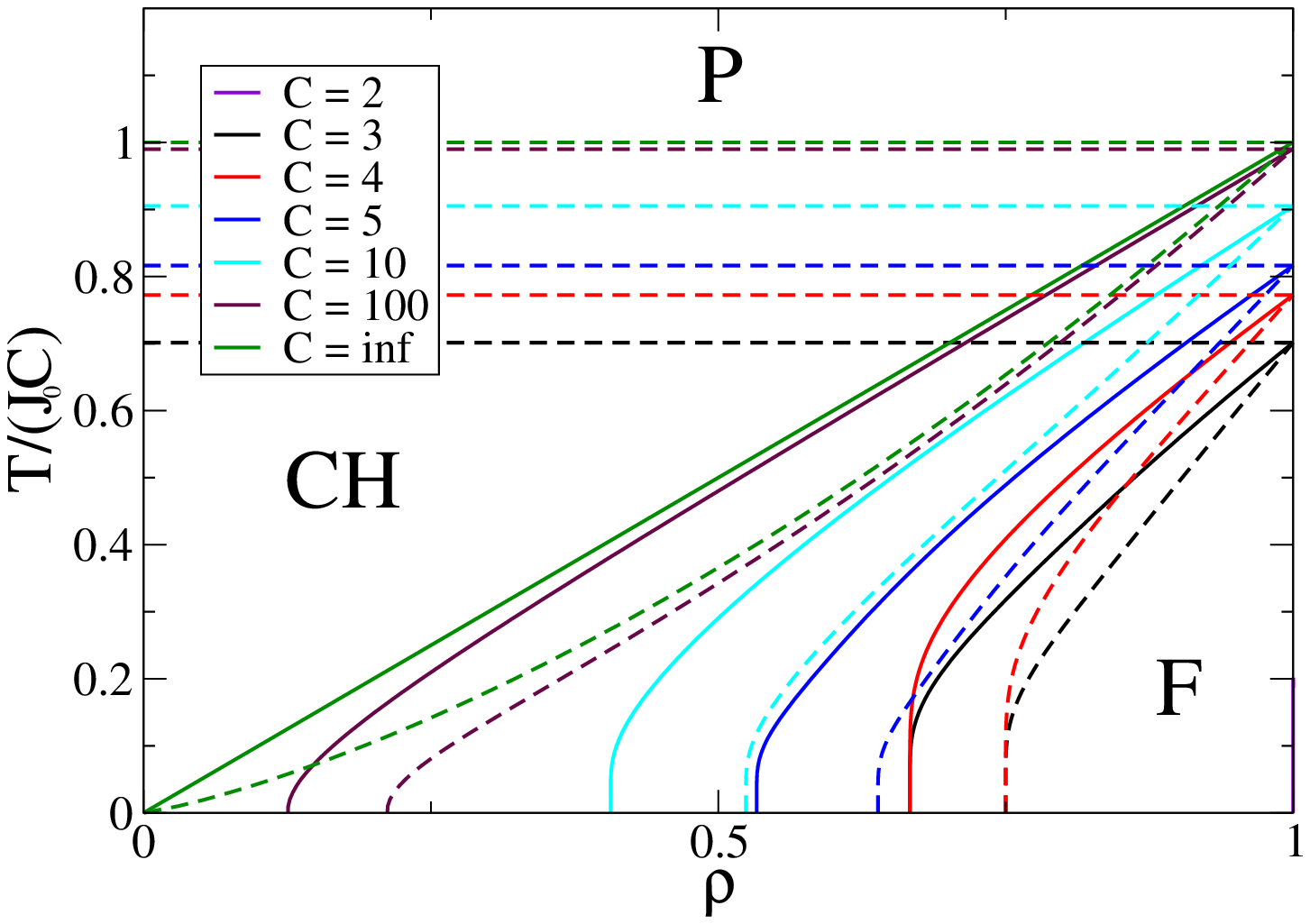}
\caption{The P to F transition lines (solid) for the Ising model on an asymmetric Bethe graph are presented as a function of the rescaled temperature $T/(J_0C)$ and the bias $\rho$ in the couplings  (see equation (\ref{eq:distriRho})).  Phase transition lines for different connectivities $C$ are shown.  The dashed lines enclose the regions where the model behaves chaotically. } \label{fig:AsBethe}
\end{center}
\end{figure}

$\\$
Using the above equations (\ref{eq:mAs}), (\ref{eq:corr}) and (\ref{eq:Wm}) we can find the stationary solutions. We consider the stationary solution $m(t+1) = m(t) = m$.   Substitution of this ansatz in (\ref{eq:mAs}) shows that $m$ is a solution of 
\begin{eqnarray}
\fl  m  = \prod_{0<\ell\leq C}\left[\sum_{\sigma_{\ell}}\left(\frac{1+\sigma_{\ell}m}{2}\right)\right]
\Bigg\langle \tanh\left[\beta\left(\sum_{0<\ell'\leq  C}J_{\ell'}\sigma_\ell\right)\right]\Bigg\rangle_{J_1,J_2,\cdots,J_{C}} \:.\label{eq:Fm}
\end{eqnarray}
  The model has a phase transition between a ferromagnetic phase (F-phase) with $m>0$ at low temperatures and a paramagnetic phase (P-phase) with $m=0$ at high temperatures.  Because this transition is continuous it is possible to determine the P to F phase transition line with an expansion of the right hand side of (\ref{eq:Fm}) around $m=0$.  The critical inverse temperature $\beta^*$ between the P-phase and F-phase is the solution of:
\begin{eqnarray}
 1 = \rho  2^{-C}\sum^{C}_{r=0}\left(\begin{array}{c} C \\r \end{array}\right)\left|2r-C\right|\tanh\left(\beta^* J \left|2r-C\right|\right) \:. \label{eq:TransFerro}
\end{eqnarray} 
Equation (\ref{eq:TransFerro}) holds for the bimodal distributions $R(J)$ of the form (\ref{eq:distriRho}).
Using the stationary ansatz $q = C(t,t') = C(t-n, t'-n)$ in (\ref{eq:corr}) we can try to find a phase transition between a paramagnetic phase with $q=0$ and a spin glass phase (SG-phase) with $q>0$.  Analogously to \cite{Hat2004} we find that $q=0$ for all temperatures and biases $\rho$.  In figure \ref{fig:AsBethe} we show the P to F transitions (solid lines) for different values of the connectivity $C$ as a function of the temperature $T$ and the bias $\rho$ in the couplings . 

\subsection{The Chaotic Phase}
Although the spin model studied in this section has no SG-phase, it has a chaotic phase (CH-phase).  In order to find this phase it is necessary to consider the dynamics of two set of spins $\bsigma$ and $\btau$ that interact on the same graph with the same thermal noise with a slightly different initial configuration.  The transition probability of the spins $(\sigma_i, \tau_i)$ is given by (\ref{eq:WDynErgo}) with $h_i(s) = \sum_{j\in \partial^{\rm in}_i}J_{ji}\sigma_j(s-1)$ and $g_i(s) = \sum_{j\in \partial^{\rm in}_i}J_{ji}\tau_j(s-1)$.  

As done in (\ref{sec:Bethe}) it is possible to derive the recursive equations for the single time marginals $\overline{P}^{\rm{d}}_t(\sigma, \tau)$.  We define the magnetisation $m_t$ and the thermal average of the Hamming distance $d_t$ between the sets $\bsigma$ and $\btau$ through
\begin{eqnarray}
 \overline{P}^{\rm{d}}_t(\sigma,\tau) = \frac{1}{4}\left(1 + m_t \sigma + m_t \tau + (1-2d_t)\sigma\tau\right) \:.
\end{eqnarray}
In the case of a bimodal distribution $R(J)$ we get for $d_t$ the recursive equation
\begin{eqnarray}
 \fl d_t &=  \sum^{C-1}_{n=0}\left(\begin{array}{c} k^{\rm in}\\ n\end{array}\right)\left(d_{t-1}\right)^{k^{\rm in}- n}\left(1-d_{t-1}\right)^n
\nonumber \\ 
\fl 
 &\int dx dy \left|\frac{\exp\left(\beta (x + |y|)\right) }{2\cosh\left(\beta (x + |y|)\right)} - \frac{\exp\left(\beta (x - |y|)\right) }{2\cosh\left(\beta (x - |y|)\right)}\right|
\nonumber \\
\fl& \sum^{n}_{v=0}\left(\begin{array}{c} n\\  v\end{array}\right)\left(\frac{1+\rho m_{t-1}/(1-d_{t-1})}{2}\right)^{v}\left(\frac{1-\rho m_{t-1}/(1-d_{t-1}) }{2}\right)^{n-v} \delta\left(x -2v + n\right) 
\nonumber \\ 
 \fl& \sum^{k^{\rm in}-n}_{w=0}\left(\begin{array}{c} k^{\rm in}-n \\  w\end{array}\right)2^{n-k^{\rm in}} \delta\left(y -2w + k^{\rm in}-n\right) 
\:. \label{eq:deDyn}
\end{eqnarray}
The time evolution of the  magnetisation $m_t$ is given by equation (\ref{eq:mAs}). In the case of a Gaussian distribution of the couplings we find for $d_t$ the equation derived in \cite{Der1987}. Starting from an initial configuration with $d_0\approx 0$, the system is said to be chaotic when the Hamming distance satisfies $d_t>0$ for large times $t\rightarrow \infty$.  In the CH-phase two paths that are initially close to each other diverge for $t\rightarrow \infty$.

 We consider a stationary ansatz $d_t = d$ and $m_t = m$ in equation (\ref{eq:deDyn}).  When $d>0$ we say that the system is chaotic.   Because the transitions are continuous we can study the bifurcations around the $d=0$  solution.  We find the following equation for the inverse transition temperature $\beta^*$ to the CH-phase
\begin{eqnarray}
 \fl 1 &=  \sum^{\infty}_{k^{\rm out}\geq 0}\sum^{\infty}_{k^{\rm in}\geq 0}k^{\rm in}\sum^{\rm{min}\it(k^{\rm out}, k^{\rm in})}_{u=0}\frac{p(k^{\rm in}, k^{\rm out},u)(k^{\rm out}-k^{\rm sym})}{c_{\rm{out}}-c_{\rm sym}}
\nonumber \\ 
\fl&
 \int dx dy \left|\frac{\exp\left(\beta^* (x + 1)\right) }{2\cosh\left(\beta^* (x + 1)\right)} - \frac{\exp\left(\beta^* (x - 1)\right) }{2\cosh\left(\beta^* (x - 1)\right)}\right|
\nonumber \\ 
\fl&\sum^{k^{\rm in}-1}_{v=0}\left(\begin{array}{c} k^{\rm in}-1\\  v\end{array}\right)\left(\frac{1+\rho m}{2}\right)^{v}\left(\frac{1-\rho m}{2}\right)^{k^{\rm in}-1-v} \delta\left(x -2v + k^{\rm in}-1\right) 
\:. \label{eq:ergodBreak}
\end{eqnarray}
For $C\rightarrow \infty$ equation (\ref{eq:ergodBreak}) reduces to $T= 4\:e^{-2\beta^*\rho m}\left(1+ e^{-2\beta^*\rho m}\right)^{-2}$.
In figure \ref{fig:AsBethe} the different phase transitions are shown.  For $\rho$ large enough, and decreasing the temperature starting from a large value we obtain subsequently  the following phases: P-phase, CH-phase, the chaotic part of the F-phase and the non-chaotic part of the F-phase.

\section{Neural Network on a Scale-Free Graph}\label{sec:Neural}
The interactions between neurons in organisms are most of the time asymmetric.  Introducing asymmetric couplings in models for neural networks  increases the biological realism of the models under study.   That is why in \cite{Cris1988, Der1987Co} the Hopfield model was generalized to include asymmetric couplings. 

We add some more realism to the model by defining the neural network on a graph with a given degree distribution.  Many real-world networks have a degree distribution of the form $p(k) = a k^{-\gamma}$, with $a$ a normalisation constant.  These are called scale-free graphs.  One example is the network of brain activity which has scale-free features \cite{Victor2005}.   In \cite{Isaac2004} neural networks on scale-free graphs with only symmetric couplings were studied.  

We consider a neural network on a fully asymmetric graph with the following distribution of indegrees and outdegrees \cite{Schwart2002}, 
\begin{eqnarray}
 p(k^{\rm in}, k^{\rm out}) = A a k^{-\lambda}_{\rm out} \delta\left(k^{\rm in}, k^{\rm out}\right) + (1-A)a^2 k^{-\lambda}_{\rm in}k^{-\lambda}_{\rm out}\:. \label{eq:degreeCorr}
\end{eqnarray}
The correlation factor $A$ in (\ref{eq:degreeCorr}) denotes the fraction of sites where the number of connections entering and leaving the site are equal.  Real-world networks have correlations between the indegrees and outdegrees \cite{Schwart2002}.  This correlation between the degrees will turn out to have much influence on the performance of a scale-free neural network.    For fixed $\lambda$ we will change the average number of interactions by increasing the lower bound $b$: $p(k^{\rm in}, k^{\rm out}) = 0$ for $k^{\rm in}<b$ and $k^{\rm out}<b$.

We take the strengths of the interactions $J_{ij}$ according to the Hebb rule:
\begin{eqnarray}
 J_{ij} = \frac{1}{p}\sum^p_{\mu=1}\xi^\mu_i\xi^\mu_j \:, \label{eq:int}
\end{eqnarray}
with the $\xi^{\mu}_i \in \left\{-1, 1\right\}$, uncorrelated patterns drawn from the probability distribution 
\begin{eqnarray}
 q(\xi^{\mu}_i) = \frac{1}{2}\left(\delta(\xi^\mu_i; 1) + \delta(\xi^\mu_i; -1)\right) \:. \label{eq:q}
\end{eqnarray}
The network has $p$ patterns $\bxi_i = \left(\xi^1_i, \xi^2_i, \cdots, \xi^p_i\right)$ on each site.  Because the $J_{ij}$ are not i.i.d.r.v. variables we can not use the equations (\ref{eq:Pout}) and (\ref{eq:PSym}).  
\subsection{Glauber Dynamics} 
We first derive the recursive equations for the marginal distributions when the variables evolve through a Glauber dynamics. From these equations we determine the phase transition from a P-phase to a retrieval phase (R-phase).  In the R-phase the network can recover a stored pattern while in the P-phase the noise is too large to retrieve a stored pattern from a distorted signal. To calculate the mean of the cavity equations (\ref{eq:DynCav}) over the quenched variables, it is necessary to define the sublattices $I_{\bxi}$: $I_{\bxi} \equiv \left\{i \in V| \bxi_i = \bxi\right\}$.   The averaged path probabilities $\overline{P}^{\rm{d}}_{\bxi}(\bsigma^{(t)})$ and  $\overline{P}^{\rm{real}}_{\bxi}(\bsigma^{(t)})$, on the sublattices  $I_{\bxi}$ are defined as
\begin{eqnarray}
 \overline{P}^{\rm{d}}_{\bxi}\left(\sigma^{0..t}\right) \equiv \frac{\sum_{i\in I_{\bxi}}\sum_{(i,\ell)\in E^{\rm{d}}}P^{(\ell)}_i\left(\sigma^{0..t}|0^{1..t}\right)}{\sum_{i\in I_{\bxi}}\sum_{(i,\ell)\in E^{\rm{d}}}}\:, \\ 
 \overline{P}^{\rm{real}}_{\bxi}(\sigma^{0..t}) \equiv \frac{\sum_{i\in I_{\bxi}}P_i(\sigma^{0..t}|0^{1..t})}{\sum_{i\in I_{\bxi}}}\:. 
\end{eqnarray}
When the graph is drawn from an ensemble defined by a degree distribution of the form $p(k^{\rm in}, k^{\rm out}, k^{\rm sym}) = p(k^{\rm in}, k^{\rm out}) \delta(k^{\rm sym})$, such that there are no symmetric couplings, we get the following recursive equation for $\overline{P}^{\rm{d}}_{\bxi}$
\begin{eqnarray}
 \fl \overline{P}^{\rm{d}}_{\bxi}\left(\sigma^{0..t}\right) = \sum^{\infty}_{k^{\rm in}=b}p(k^{\rm in}) \frac{c_{\rm out}\left(k^{\rm in}\right)}{c_{\rm{out}}} \prod_{0<\ell\leq k^{\rm in}}\sum_{\sigma^{0..t-1}_{\ell}} \sum_{\bxi_{\ell}}\frac{\overline{P}^{\rm{d}}_{\bxi_\ell}(\sigma^{0..t-1}_{\ell})}{2^p} 
\nonumber \\ 
 p_0(\sigma(0))\prod_{s\geq 0}\frac{\exp\left[\beta \sigma(s+1) \sum_{0<\ell'\leq k^{\rm in}}\frac{\bxi\cdot\bxi_\ell'}{p}\sigma_{\ell'}(s)\right]}{2\cosh\left[\beta \sum_{0<\ell'\leq k^{\rm in}}\frac{\bxi\cdot\bxi_\ell'}{p}\sigma_{\ell'}(s)\right]}\:.  \label{eq:PoutNeur}
\end{eqnarray}
In the above the $c_{\rm out}(k^{\rm in})$ is the average number of directed edges leaving a site, given the indegree $k^{\rm in}$: $c_{\rm out}(k^{\rm in}) = \sum_{k^{\rm out}}p(k^{\rm out}|k^{\rm in})k^{\rm out}$.
For the averaged path probability on the original graph we obtain analogously
\begin{eqnarray}
 \fl \overline{P}^{\rm{real}}_{\bxi}\left(\sigma^{0..t}\right)  =\sum^{\infty}_{k^{\rm in}=b}p(k^{\rm in}) \prod_{0<\ell\leq k^{\rm in}} \sum_{\sigma^{0..t-1}_{\ell}}\sum_{\bxi_{\ell}}\frac{\overline{P}^{\rm{d}}_{\bxi_\ell}(\sigma^{0..t-1}_{\ell})}{2^p}
\nonumber\\
 p_0(\sigma(0))\prod_{s\geq 0}\frac{\exp\left[\beta \sigma(s+1) \sum_{0<\ell'\leq k^{\rm in}}\frac{\bxi\cdot\bxi_\ell}{p}\sigma_{\ell'}(s)\right]}{2\cosh\left[\beta\sum_{0<\ell'\leq k^{\rm in}}\frac{\bxi\cdot\bxi_{\ell'}}{p}\sigma_{\ell'}(s)\right]} \:.\label{eq:PRealNeur}
\end{eqnarray}
We will use the notation $P^{\rm a}_{\bxi}(\sigma(t)) = \frac{1}{2}\left(1+m^{\rm a}_{\bxi}(t)\sigma(t)\right)$ with superscript $\rm{a} = \rm{d}$ or $\rm{a} = \rm{real}$.  The magnetisations $m^{\rm d}_{\bxi}$ and $m^{\rm real}_{\bxi}$ evolve in time according to
\begin{eqnarray}
  \fl m^{\rm a}_{\bxi}(t) = \sum^{\infty}_{k^{\rm in} = b}p(k^{\rm in})\frac{c_{\rm a}\left(k^{\rm in}\right)}{c_{\rm{out}}}  2^{-pk^{\rm in}}\prod_{0<\ell\leq k^{\rm in}}\sum_{\sigma_\ell}\sum_{\bxi_\ell}\left(\frac{1+\sigma_\ell m^{\rm a}_{\bxi_\ell}(t-1)}{2}\right)  \nonumber \\ 
\tanh\left(\beta \frac{\sum_{0<\ell'\leq k^{\rm in}}\bxi\cdot \bxi_{\ell'}\sigma_{\ell'}}{p}\right) \label{eq:moutV}\:,
\end{eqnarray}
with the average connectivities
\begin{eqnarray}
c_{\rm{d}}(k^{\rm in}) = c_{\rm {out}}(k^{\rm in}) = (1-A)c_{\rm out} + A\:k^{\rm in}  \:,\\ 
c_{\rm{real}}(k^{\rm in}) = c_{\rm out} \:.
\end{eqnarray}
We simplify the equations with the condensed ansatz $m^{\rm a}_{\bxi}(t) = \xi^1m^{\rm a}(t)$.  This ansatz assumes that the spins (= neurons) only have a finite overlap with the first pattern.   The overlap $m^{\rm a}(t)$ evolves according to
\begin{eqnarray}
  \fl m^{\rm a}(t)  = \sum^{\infty}_{k^{\rm in}=b}p(k^{\rm in})\frac{c_{\rm a}(k^{\rm in})}{c_{\rm{out}}}M(k^{\rm in})\:, \label{eq:NNmOut}
\end{eqnarray}
with
\begin{eqnarray}
\fl M(k^{\rm in}) =  2^{-(p-1)k^{\rm in}}\sum^{k^{\rm in}(p-1)}_{r=0}\sum^{k^{\rm in}}_{s=0}\left(\begin{array}{c} k^{\rm in}(p-1)\\r \end{array}\right)\left(\begin{array}{c} k^{\rm in}\\s \end{array}\right) 
\nonumber \\ 
\left[\frac{1+m^{\rm{d}}(t-1)}{2}\right]^s\left[\frac{1-m^{\rm{d}}(t-1)}{2}\right]^{k^{\rm in}-s}\tanh\left(\frac{\beta(2s+2r-k^{\rm in}p)}{p}\right) \:. \nonumber \\
\end{eqnarray}
When calculating numerically the sum in the degrees $k^{\rm in}$ in equation (\ref{eq:NNmOut}) we have to introduce a cutoff $K$.  We will bound $m^{\rm a}$ by two values, $m^{\rm a}_{\rm{l}}<m^{\rm a}$ and $m^{\rm a}_{\rm{u}}>m^{\rm a}$, with $m^{\rm a}_{\rm{l}}$ and $m^{\rm a}_{\rm{u}}$ defined through
\begin{eqnarray}
\fl m^{\rm a}_{\rm{l}}(t) \equiv \sum^K_{k^{\rm in} = b}p(k^{\rm in})\frac{c_{\rm a}(k^{\rm in})}{c_{\rm{out}}}M(k^{\rm in}) \:,
\label{eq:ml} \\ 
\fl m^{\rm a}_{\rm{u}}(t) \equiv \sum^K_{k^{\rm in} = b}p(k^{\rm in})\frac{c_{\rm a}(k^{\rm in})}{c_{\rm{out}}}M(k^{\rm in}) 
 +\rm{sign}\it\left(m^{\rm d}_{\rm u}(t-1)\right)\sum^{\infty}_{k^{\rm in} \it = K+1}\it p(k^{\rm in}) \frac{c_{\rm a}(k^{\rm in})}{c_{\rm{out}}} \label{eq:mu} \:.
\end{eqnarray}
In equation (\ref{eq:mu}) we used that $M(\infty) = \rm {sign}\left(\it m^{\rm d}\it(t-1)\right)$. Because we have a power-law decay of the degree distribution (and not an exponential decay) it is important to take the cutoff $K$ into consideration when we want to know the asymptotic behaviour of the neural network for $K\rightarrow \infty$.  The macroscopic observables will converge much slower to the asymptotic value $K=\infty$ when the degree distribution is power law.   For finite $K$ the time evolution of equation (\ref{eq:ml}) is confirmed by Monte Carlo simulations. 

\subsection{The Retrieval State}
\begin{figure}[h!]
\begin{center}
\includegraphics[angle=-90, width=.6 \textwidth]{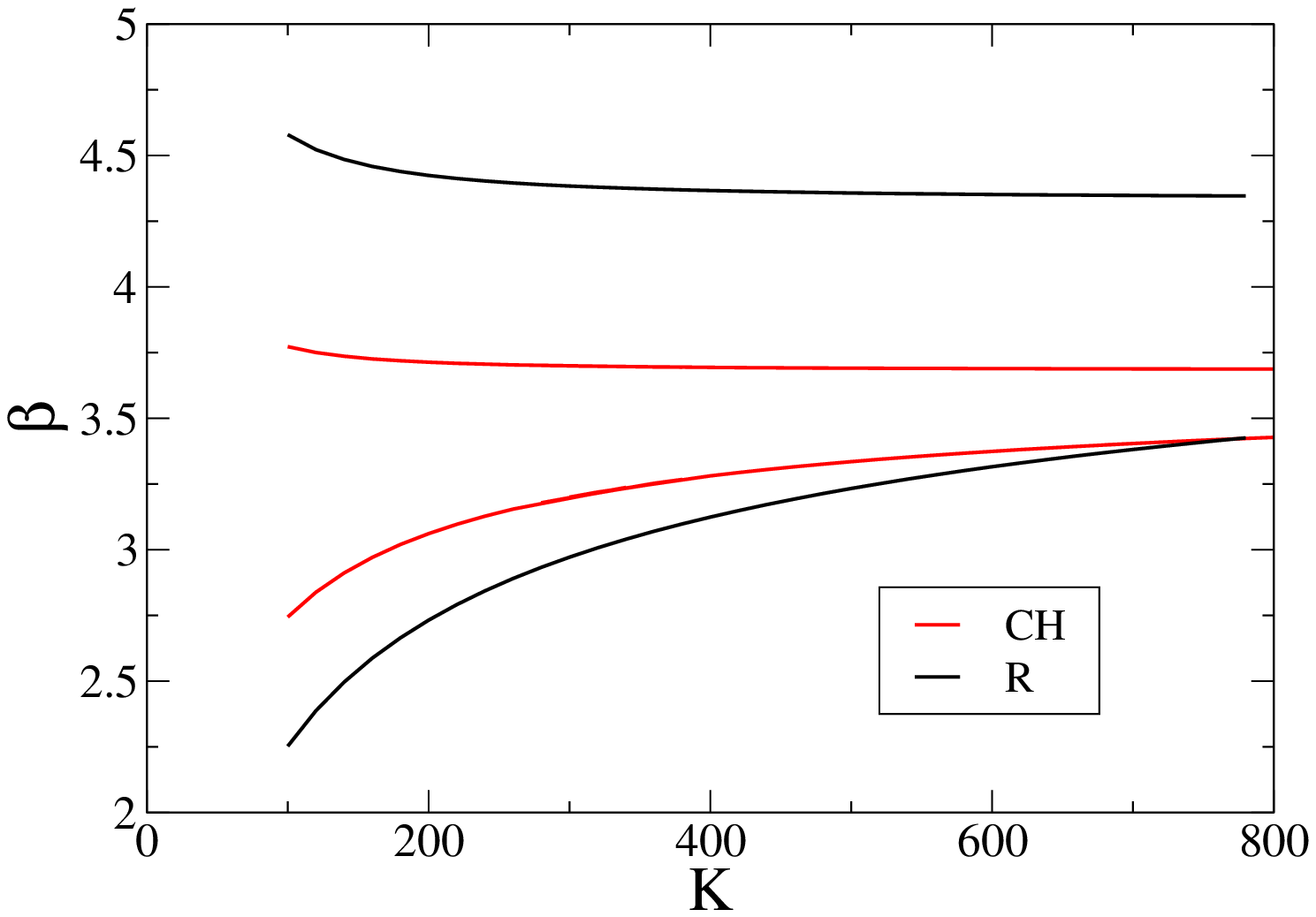}
\caption{Neural network on a scale-free graphs: The bounds $\beta^{\rm{R}}_{\rm{u}}, \beta^{\rm{ch}}_{\rm{u}}$ (upper lines) and $\beta^{\rm{R}}_{\rm{l}}, \beta^{\rm{ch}}_{\rm{l}}$ (lower lines) on the inverse critical temperatures $\beta^{\rm{R}}, \beta^{\rm{ch}}$ as a function of the cutoff $K$.  The bounds on $\beta^{\rm{R}}$ are calculated for the model parameters $\lambda=2$, $p=3$, $b=4$.  The bounds on $\beta^{\rm{ch}}$ are calculated for $\lambda=2.2$, $p=3$, $b=4$. The upper bounds saturate much faster than the lower bounds.} \label{fig:conv}
\end{center}
\end{figure}

\begin{figure}[h!]
\begin{center}
\includegraphics[angle=-90, width=.6 \textwidth]{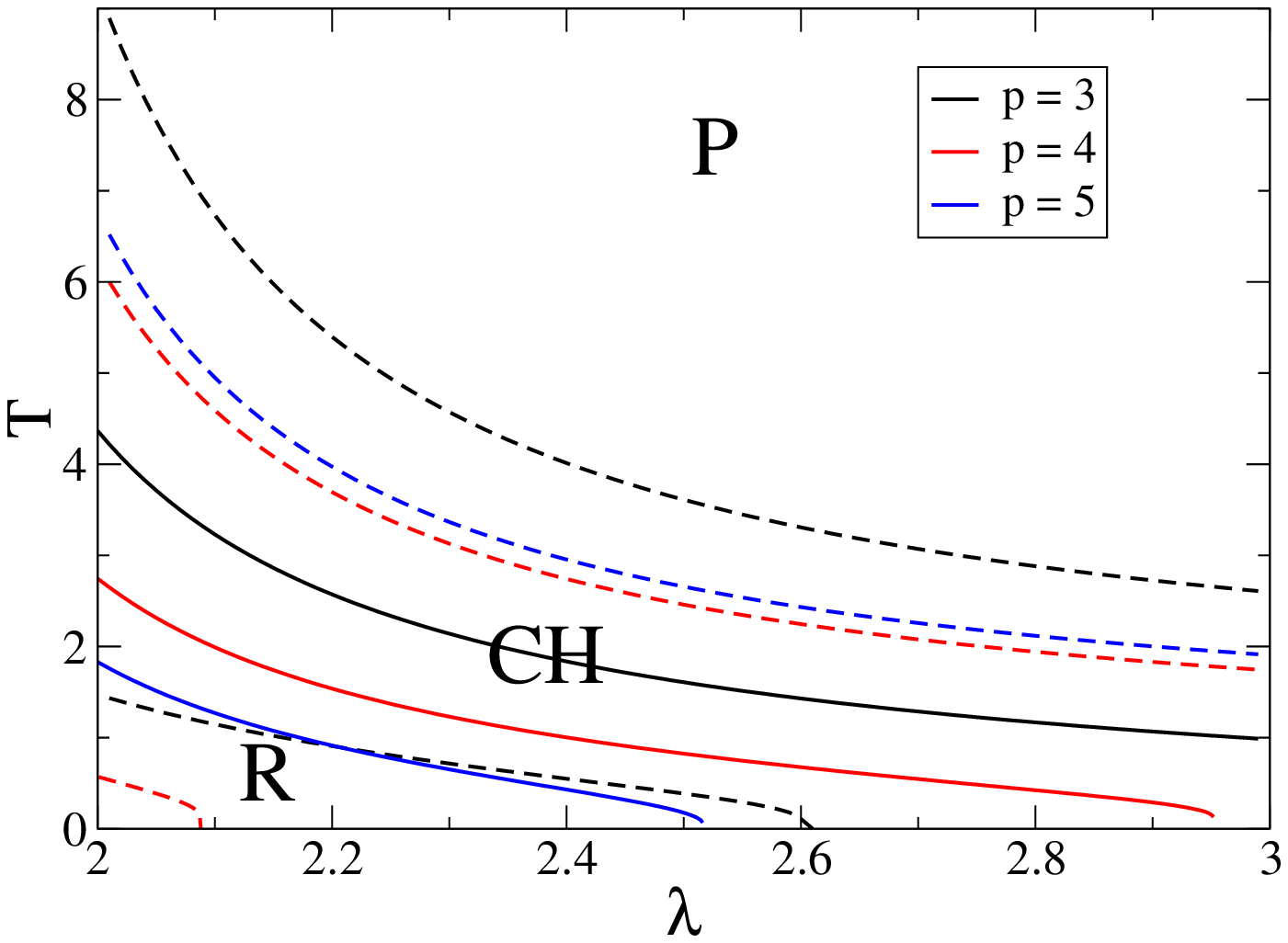}
\caption{Neural network on a scale-free graphs: 
The critical temperatures $T^{\rm{R}}$ (solid lines), $T^{\rm{ch}}$ (upper dashed lines) and $T^{\rm{m}}$ (lower dashed lines) as a function of the exponent $\lambda$ (see equation (\ref{eq:degreeCorr})) for a different number of patterns $p$. The minimal indegree is $b=4$ and the correlation factor $A=0$.  The R-phase is located below the solid line.  The neural network is chaotic between the dashed lines.} \label{fig:gammaVar}
\end{center}
\end{figure}

\begin{figure}[h!]
\begin{center}
\includegraphics[angle=-90, width=.6 \textwidth]{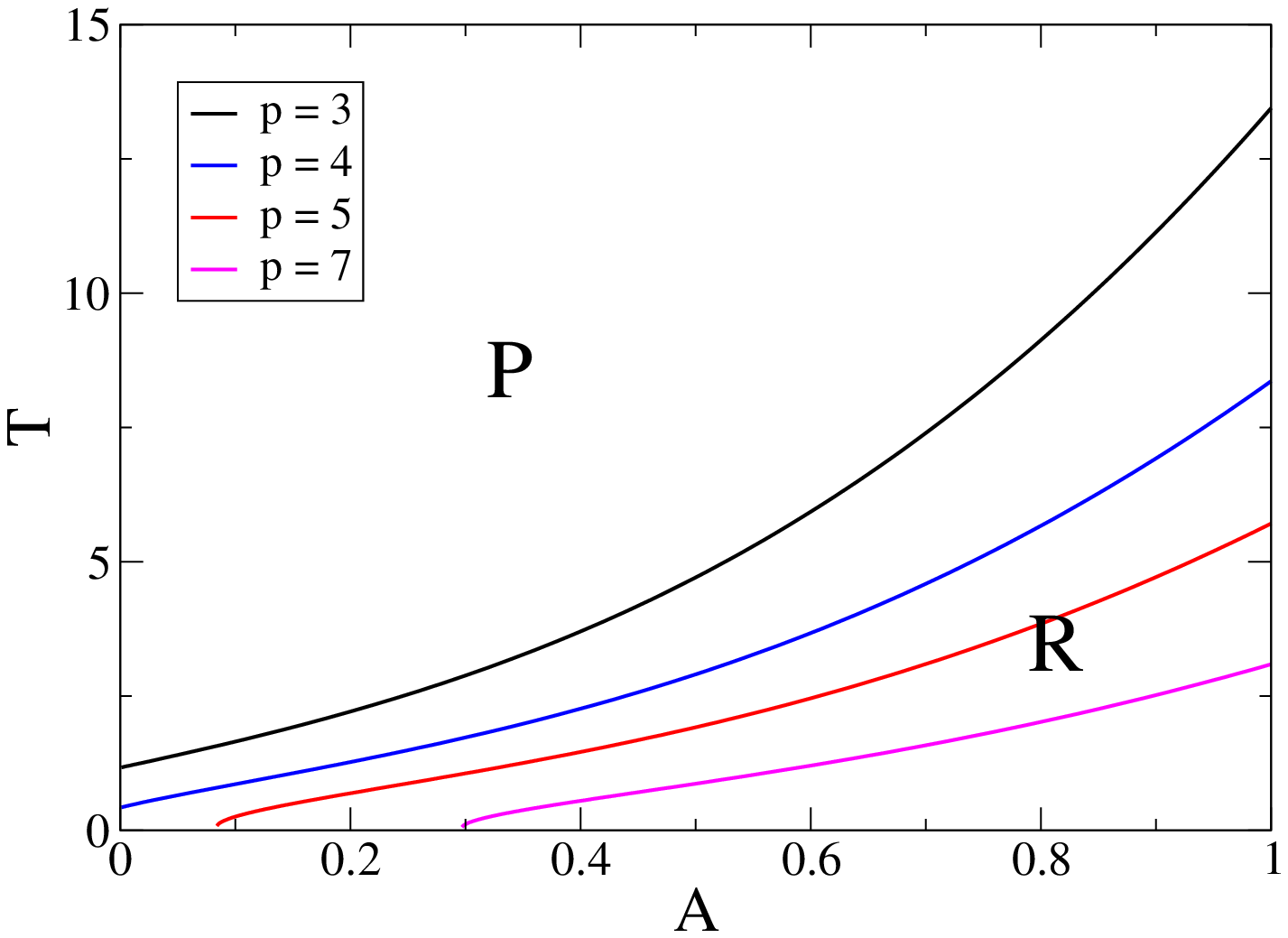}
\caption{The critical temperatures of the retrieval state $T^{\rm{R}}$ of the neural network as a function of the correlation factor $A$  of the scale-free graph for a different number of patters $p$ are compared. The ensemble of scale-free graphs has the parameters $\lambda = 2.8$, $b=4$.  The critical temperatures are estimated with $T^{\rm{R}}_{\rm u}$ for $K = 2000$.  The R-phase increases considerably with the correlation factor $A$.} \label{fig:corr2.5}
\end{center}
\end{figure}

When we consider a stationary state of the form, $m(t) = m(t-1)$, we get the following equation for the critical inverse temperature $\beta^{\rm{R}}$ of the P to R transition
\begin{eqnarray}
 \fl 1 =  \sum_{k^{\rm in}}p(k^{\rm in})\frac{c\left(k^{\rm in}\right)}{c_{\rm{out}}} \mathcal{A}(\beta^{\rm{R}}, k^{\rm{in}}) \:, \label{eq:RP}
\end{eqnarray}
with
\begin{eqnarray}
 \fl \mathcal{A}(\beta^{\rm{R}}, k^{\rm{in}}) =  2^{-pk^{\rm in}}\sum^{k^{\rm in}(p-1)}_{r=0}\sum^{k^{\rm in}}_{s=0}\left(\begin{array}{c} k^{\rm in}(p-1)\\r \end{array}\right)\left(\begin{array}{c} k^{\rm in}\\s \end{array}\right)
\nonumber \\
 (2s-k^{\rm in})\tanh\left(\frac{\beta^{\rm{R}}(2s+2r-k^{\rm in}p)}{p}\right)  \:.  
\end{eqnarray} 
Below the critical temperature $T_R = 1/\beta^{R}$, $m^{\rm real}>0$ such that the neural network can retrieve a stored pattern from a distorted initial configuration.  To calculate numerically equation (\ref{eq:RP}) we take a distribution $p(k^{\rm in}) = ak^{-\gamma}_{\rm in}$ with $k^{\rm in}\in [b, \cdots, K]$ and zero for other values of $k^{\rm in}$.   We introduce the lower critical value $\beta^{\rm{R}}_{\rm{l}}$, and  the upper value $\beta^{\rm{R}}_{\rm{u}}$ through
\begin{eqnarray}
\fl 1 =\sum^K_{k^{\rm in}=b}p(k^{\rm in})\frac{c\left(k^{\rm in}\right)}{c_{\rm{out}}}\mathcal{A}(\beta^{\rm{R}}_{\rm{l}}, k^{\rm in}) \:,
\nonumber \\ 
 \fl 1 =\sum^K_{k^{\rm in}=b}p(k^{\rm in})\frac{c\left(k^{\rm in}\right)}{c_{\rm{out}}}\mathcal{A}(\beta^{\rm{R}}_{\rm{u}}, k^{\rm in}) + \sqrt{\frac{2}{\pi p}}\left(\sum^{\infty}_{k^{\rm in}=K+1}p(k^{\rm in})\sqrt{k^{\rm in}}\frac{c\left(k^{\rm in}\right)}{c_{\rm{out}}}\right)\:.  \label{eq:critUpRetr}
\end{eqnarray}
To derive equation (\ref{eq:critUpRetr}) we used that $\mathcal{A}(\beta, C)$ has the following asymptotic behaviour for $C\rightarrow \infty$
\begin{eqnarray}
 \mathcal{A}(\beta, C) \rightarrow \sqrt{\frac{2}{\pi p}}\sqrt{C} \:. 
\end{eqnarray}    
The asymptotic value of the Riemann sum in the second term of (\ref{eq:critUpRetr}) can be calculated using a series that converges exponentially in $K$ \cite{Bor2000}.   In figure \ref{fig:conv} we compare how $\beta^{\rm{R}}_{\rm{l}}$ and  $\beta^{\rm{R}}_{\rm{u}}$ converge to their asymptotic value for $K\rightarrow \infty$. The upper bound $\beta^{\rm{R}}_{\rm{u}}$ clearly converges fast to the asymptotic value.  The lower bound $\beta^{\rm{R}}_{\rm{l}}$ on the other hand converges very slow to its asymptotic value.  When we would have bounded the critical value $\beta^{\rm{R}}$ from below with $\beta^{\rm{R}}_{\rm{l}}$, which we do usually for Poissonian graphs, we would have obtained a bad estimate of the critical temperature $\beta^{\rm{R}}$.  Therefore, we estimate in figures \ref{fig:gammaVar} and \ref{fig:corr2.5} the critical temperatures $T^{\rm R}$ with the upper bounds $T^R_{\rm{u}}$.  There we plotted the critical temperatures $T^{\rm{R}}$ as a function of, respectively, the exponent $\lambda$ and the correlation factor $A$.  The retrieval phase increases with the exponent $\lambda$, which is expected because the mean connectivity of the graph also increases with $\lambda$.  Increasing the correlation $A$ between the indegrees and the outdegrees on scale-free graphs has a positive effect on the performance of the network.  The neural network becomes much more tolerant to noise and can retrieve considerably more patterns when $A$ increases. 

\subsection{The Chaotic Phase}
In this subsection we determine the CH-phase of the neural network. In \cite{Der1987Co} this was done on a Poissonian graph using annealed methods.
We consider two systems on the same graph undergoing the same thermal noise through the coupled dynamics of \ref{subsec:coupled}.  We find the following Markovian process for the single time marginals $\overline{P}^{\rm d}_{t, \bxi}$:
\begin{eqnarray}
\fl \overline{P}^{\rm d}_{t, \bxi}\left(\sigma, \tau\right) = \sum_{k^{\rm in}}p(k^{\rm in})\frac{c(k^{\rm in})}{c_{\rm out}}\prod_{0< \ell \leq k^{\rm in}}\sum_{\sigma_\ell, \tau_\ell}\sum_{\bxi_{\ell}}\frac{\overline{P}^{\rm d}_{t-1, \bxi_\ell}\left(\sigma_\ell, \tau_\ell\right)}{2^{p}} 
\nonumber \\ 
 W_{\rm{c}}\left[\sigma, \tau|\sum_{0< \ell' \leq k^{\rm in}}\frac{\bxi\cdot \bxi_{\ell'}}{p}\sigma_{\ell'},\sum_{0< \ell' \leq k^{\rm in}}\frac{\bxi\cdot \bxi_{\ell'}}{p}\tau_{\ell'} \right]\:, \label{eq:PChdyn}
\end{eqnarray}
where $W_{\rm{c}}$ is the transition probability defined in (\ref{eq:WDynErgo}).
We parametrise the single time marginals $\overline{P}^{\rm d}_{t, \bxi}$ with the magnetisations $m_{\bxi, 1}(t)$, $m_{\bxi, 2}(t)$ and the Hamming distance $d_{\bxi}(t)$ at time step $t$:
\begin{eqnarray}
\overline{P}^{\rm d}_{t, \bxi}\left(\sigma, \tau\right) = \frac{1}{4}\left[1 + \sigma\: m_{\bxi, 1}(t) + \tau\: m_{\bxi, 2}(t) + \sigma\tau \left(1-2 d_{\bxi}(t)\right)\right] \:.
\end{eqnarray}
The Hamming distance evolves according to
\begin{eqnarray}
\fl d_{\bxi}(t) &=  \sum^{\infty}_{k^{\rm in}=b}p(k^{\rm in})
\nonumber \\
\fl 
&\prod_{0<\ell\leq k^{\rm in}}\sum_{\sigma_\ell, \tau_\ell}\sum_{\bxi_\ell} 2^{-p}\left(\frac{1+\sigma_\ell m_{\bxi_\ell, 1}(t-1)+ \tau_\ell m_{\bxi_\ell, 2}(t-1) + \left(1-2d_{\bxi_\ell}(t-1)\right)\sigma_\ell \tau_\ell}{4}\right)
\nonumber \\
\fl &\left|\frac{\exp\left[\beta \sum_{\ell'} \left(\frac{\bxi\cdot\bxi_{\ell'}}{p}\right)\sigma_{\ell'}\right]}{2\cosh\left[\beta \sum_{\ell'} \left(\frac{\bxi\cdot\bxi_{\ell'}}{p}\right)\sigma_{\ell'}\right]} - \frac{\exp\left[\beta \sum_{\ell'} \left(\frac{\bxi\cdot\bxi_{\ell'}}{p}\right)\tau_{\ell'}\right]}{2\cosh\left]\beta \sum_{\ell'} \left(\frac{\bxi\cdot\bxi_{\ell'}}{p}\tau_{\ell'}\right)\right]} \right|\:.
\end{eqnarray}
The magnetisations $m_{\bxi, 1}(t)$  and  $m_{\bxi, 2}(t)$ behave according to equation (\ref{eq:moutV}).
We use the condensed initial conditions, $m_{\bxi, 1}(0) =\xi^1m(0)$, $m_{\bxi, 2}(0) = \xi^1m(0)$ and $d_{\bxi}(0) = d(0)$.  These initial conditions imply that the initial configurations have a finite overlap with only the first pattern.  From the time evolution (\ref{eq:PChdyn}) we find that for the condensed initial conditions $m_{\bxi, 1}(t) =\xi^1m(t)$, $m_{\bxi, 2}(t) = \xi^1m(t)$ and $d_{\bxi}(t) = d(t)$.  The  evolution of the overlap $m(t)$ is given by equation (\ref{eq:NNmOut}).  For $d(t)$ we get 
\begin{eqnarray}
 \fl d(t)  &=  \sum^{\infty}_{k^{\rm in}=b}p(k^{\rm in})\sum^{k^{\rm in}}_{n=0} \left(\begin{array}{c} k^{\rm in}\\ n\end{array}\right) \left(d(t-1)\right)^{k^{\rm in}-n}\left(1- d(t-1)\right)^{n}
\nonumber \\ 
\fl &\sum^{(k^{\rm in}-n)(p-1)}_{R_1 = 0}\sum^{n(p-1)}_{R_2 = 0} f\left(R_1, (p-1)(k^{\rm in}-n)\right) f\left(R_2, (p-1)n\right) 
\nonumber \\ 
\fl &\int dx dy\ \exp\left[ \frac{\beta}{p}\left(- n(p-1) + 2R_2 + x\right)\right]
\nonumber \\ 
\fl 
&\left|\frac{\exp\left[\frac{\beta}{p}\left(|y| + 2R_1 -(k^{\rm in}-n)(p-1)\right) \right] }{2\cosh\left[\frac{\beta}{p} \left(x+|y|- k^{\rm in}(p-1) + 2 R_1 + 2 R_2 \right)\right]} \right.
\nonumber \\ 
\fl &\left. - \frac{\exp\left[\frac{\beta}{p}\left(-|y|- 2R_1 +(k^{\rm in}-n)(p-1)\right)\right]}{2\cosh\left[\frac{\beta}{p} \left(x - |y| + (k^{\rm in}-2n)(p-1)- 2 R_1 + 2 R_2\right)\right]}\right| 
\nonumber \\ 
\fl &
\sum^{n}_{v=0}\left(\begin{array}{c} n\\  v\end{array}\right)\left(\frac{1+ m(t-1)/(1-d(t-1))}{2}\right)^{v}
\nonumber \\ 
\fl 
&\left(\frac{1-m(t-1)/(1-d(t-1))}{2}\right)^{n-v} \delta\left(x -2v + n\right) 
\nonumber \\ 
 \fl &\sum^{k^{\rm in}-n}_{w=0}\left(\begin{array}{c} k^{\rm in}-n \\  w\end{array}\right) 2^{-n+k^{\rm in}} \delta\left(y -2w + k^{\rm in}-n\right) 
\:. \label{eq:dtEvolve}  
\end{eqnarray}
In equation (\ref{eq:dtEvolve}) we summed subsequently over the following variables: the indegrees $k^{\rm in}$,  the number of neighbouring spins $n$ with $\sigma\neq \tau$, the number of neighbouring spins $v$ with $\sigma=\tau$ and $\sigma=1$ and the number of neighbouring spins $w$ with  $\sigma \neq \tau$ and $\sigma=1$. The summation variables $R_1$ and $R_2$ are the number of non-condensed patterns $\xi^{\mu}_\ell$ on the neighbouring spins with, respectively, $\sigma\neq \tau$ and $\sigma=\tau$ that are equal to the corresponding pattern $\xi^{\mu}$ on the original site. 
The complex function $f(R; x)$ used in equation (\ref{eq:dtEvolve}) equals
\begin{eqnarray}
\fl  f(R; x) = \frac{1}{2\pi}\int^{2\pi}_0 d\omega \exp\left[i\omega R\right] \left|\cos\left(\frac{\omega}{2} \right)\right|^{x}
\exp\left[-ix\: \atan\left(\frac{\sin \omega }{1 + \cos \omega }\right)\right]\:.
\end{eqnarray}
It is possible to find an equation for the transition temperature $\beta^{\rm{ch}}$ to a CH-phase with the stationary value $d>0$ of $d(t)$ by expanding the left side of (\ref{eq:dtEvolve}) around $d(t-1)=0$:
\begin{eqnarray}
 1 = \sum_{k^{\rm in}}p(k^{\rm in})k^{\rm in}\frac{c(k^{\rm in})}{c_{\rm out}}\mathcal{B}(\beta, k^{\rm in}, m)\:, \label{eq:critd}
\end{eqnarray}
with
\begin{eqnarray}
\fl  \mathcal{B}(\beta, k^{\rm in}, m) =\int dx  \sum^{k^{\rm in}-1}_{v=0}\left(\begin{array}{c} k^{\rm in}-1\\  v\end{array}\right)\left(\frac{1+ m}{2}\right)^{v}
\left(\frac{1- m}{2}\right)^{k^{\rm in}-1-v} \delta\left(x -2v + k^{\rm in}-1\right) 
\nonumber \\ 
 \fl \sum^{p-1}_{R_1 = 0}\sum^{(k^{\rm in}-1)(p-1)}_{R_2 = 0} f\left(R_1, p-1\right) f\left(R_2, (p-1)(k^{\rm in}-1)\right) 
\nonumber \\ 
\fl 
\exp\left[\frac{\beta}{p}\left(- (k^{\rm in}-1)(p-1) + 2R_2 + x\right)\right]
\nonumber \\ 
\fl 
\left|\frac{\exp\left[\frac{\beta}{p}\left(1 + 2R_1 - (p-1)\right)\right] }{2\cosh\left[\frac{\beta}{p}\left( (x+1)- k^{\rm in}(p-1) + 2 R_1 + 2 R_2 \right)\right]}\right.
\nonumber \\
\left.
 - \frac{\exp\left[\frac{\beta}{p}\left(-1- 2 R_1 +(p-1)\right)\right]}{2\cosh\left[\frac{\beta}{p} \left((x - 1) + (-k^{\rm in}+2)(p-1)- 2 R_1 + 2 R_2\right)\right]}\right|  \:.
\end{eqnarray}
The asymptotic behaviour of $\mathcal{B}$ for $C\rightarrow \infty$ is given by: 
\begin{eqnarray}
 \mathcal{B}(\beta, C, m) \sim \exp\left(C\Psi(m)\right) \:,
\end{eqnarray}
with
\begin{eqnarray}
 \fl \Psi(m) =  \log\left[\frac{(1-m) + (1+m)\exp\left(\frac{2x}{p}\right)}{2}\right]
+(p-1)\log\left[\frac{1+\exp\left(\frac{2x}{p}\right)}{2}\right] - x \:,
\end{eqnarray}
and 
\begin{eqnarray}
 x = \frac{p}{2} \log\left[\frac{-\frac{2m}{p} + m + \sqrt{\left(-m + \frac{2m}{p}\right)^2 - (m-1)(1+m)}}{1+m}\right] \:.
\end{eqnarray}
When $|m|>0$ we have that $\Psi <0$, hence, the series converges exponentially for large $C$.  When $m=0$ we have that $\Psi= 0$.  The asymptotic behaviour of $\mathcal{B}$ is then: 
\begin{eqnarray}
\fl  \mathcal{B}(\beta, C, m) \sim \frac{1}{\sqrt{C}}\left(\sqrt{\frac{2}{p\pi}}\right) 2^{-p+1} \sum^{p-1}_{r=0}\left(\begin{array}{c} p-1 \\ r \end{array}\right) \left|1 + 2r - (p-1)\right| = \frac{\zeta}{\sqrt{C}} 
 \:.
\end{eqnarray}
We define the upper bounds $\beta^{\rm{ch}}_{\rm{u}}$ and $\beta^{\rm{ch}}_{\rm{l}}$ on the inverse critical temperature $\beta^{\rm ch}$ to the CH-phase for $m=0$ as:
\begin{eqnarray}
 1 = \sum^K_{k^{\rm in}= b }p(k^{\rm in})k^{\rm in}\frac{c(k^{\rm in})}{c_{\rm out}} \mathcal{B}(\beta^{\rm ch}_{\rm{l}}, k^{\rm in}, 0)\:,\\ 
 1 = \sum^K_{k^{\rm in}= b }p(k^{\rm in})k^{\rm in}\frac{c(k^{\rm in})}{c_{\rm out}}\mathcal{B}(\beta^{\rm ch}_{\rm{u}}, k^{\rm in}, 0) + \zeta\sum^{\infty}_{k^{\rm in}=  K+1 }p(k^{\rm in})\sqrt{k^{\rm in}}\frac{c(k^{\rm in})}{c_{\rm out}}\:.
\end{eqnarray} 
The convergence of $\beta^{\rm ch}_{\rm{u}}$ and $\beta^{\rm ch}_{\rm{l}}$ to their asymptotic value $\beta^{\rm ch}$ in function of $K$ is plotted in figure \ref{fig:conv}.  Because $\beta^{\rm ch}_{\rm{u}}$ saturates much faster we used this  value in  figure \ref{fig:gammaVar}  to estimate $\beta^{\rm ch}$.  The R-phase contains a part with $d>0$.  The bounds $\beta^{\rm m}_{\rm{u}}$ and $\beta^{\rm m}_{\rm{l}}$ on the inverse critical temperature $\beta^{\rm m}$ of the transition from the chaotic part of the R-phase to the non chaotic part of the R-phase  are calculated with substitution of respectively $m^{\rm a}_{\rm{l}}$ and $m^{\rm a}_{\rm u}$ into equation (\ref{eq:critd}).   The value of $\beta^{\rm m}_{\rm{l}}$ is a good approximation to $\beta^{\rm m}$.
In figure \ref{fig:gammaVar} the complete phase diagram of the neural network with the P-phase, the R-phase and the CH-phase is presented. The chaotic region of the neural network is enclosed by the dashed lines.  This region is larger for odd values of $p$. The R-phase and CH-phase become smaller when $p$ increases and the non chaotic part of the R-phase dissapears when $\gamma$ increases.   The CH-phase is larger for odd values of $p$ than for even values of~$p$. 
\section{The Stationary Solution}\label{sec:9}
In this section we develop an algorithm  to calculate the marginals of the stationary solution.  This algorithm would be the equivalent of the belief propagation equations (\ref{eq:CavFinal}).
Equations (\ref{eq:DynCav}) constitute an algorithm with a linear computational complexity $\mathcal{O}(N)$.  But because the algorithm scales as $\mathcal{O}(2^t)$ in time $t$ we can only follow the dynamics for a small number of time steps.  We are interested in the marginals of the stationary state when $t\rightarrow \infty$.  To solve this we introduce some assumptions on the distribution $P(\sigma^{0..t})$.  

\subsection{The One-time Approximation}
The first simple approximation one can make is to neglect the correlations in time:
\begin{eqnarray}
 P_\ell(\sigma^{0..t}|\theta^{0..t}) = \prod^t_{s=1}P^*_{\ell}(\sigma(s)|\theta(s)) = \prod^t_{s=1}\frac{\exp\left[\beta u_\ell(\theta(s))\sigma(s)\right]}{2\cosh\left[\beta u_\ell\left(\theta(s)\right)\right]}\:. \label{eq:PDet1TT}
\end{eqnarray}
 In general (see for example equations (\ref{eq:distriout}), (\ref{eq:distrisym}) and (\ref{eq:distrireal})), when solving the dynamics, we have equations of the following type 
 \begin{eqnarray}
 P(\sigma^{0..t}|\theta^{1..t}) = \mathcal{F}\left(\sigma^{0..t}|\theta^{1..t};  \left\{ P_\ell, J_\ell\right\}_{\ell = 1.. k}\right) \:. \label{eq:PDet1T}
\end{eqnarray}
We close these equations with (\ref{eq:PDet1TT}) using 
\begin{eqnarray}
P^*(\sigma(t)|\theta) = \lim_{s\rightarrow -\infty}\sum_{\sigma^{s..t-1}} P(\sigma^{s..t}|\theta^{s+1..t})\:, \label{eq:closed}
\end{eqnarray}
with $\theta^{s..t}$ the constant vector with components $\theta$.
When we insert equation (\ref{eq:PDet1TT}) in the left hand side of (\ref{eq:closed}) we find 
\begin{eqnarray}
 P^*(\sigma(t)|\theta) =  \lim_{s\rightarrow -\infty}\sum_{\sigma^{s..t-1}}\mathcal{F}\left(\sigma^{s..t}|\theta^{s+1..t};  \left\{ P^*_\ell, J_\ell\right\}_{\ell = 1..k}\right) \:. \label{eq:resultingA}
\end{eqnarray}
This approximation leads to the correct stationary solution in the case of models defined on fully symmetric or fully asymmetric graphs.  This makes us curious to see how good this approximation would work for models defined on partially asymmetric graphs.  Because we neglected correlations in time we can not expect a good description of the spin glass phase.

\subsection{Density Evolution Equations in the One-time Approximation}
We apply the approximation given by (\ref{eq:PDet1TT}) and (\ref{eq:closed}) to the equations (\ref{eq:distriout}), (\ref{eq:distrisym}) and (\ref{eq:distrireal}):
 \begin{eqnarray}
\fl  P^{\rm sym}(\sigma^{0..t}|\theta^{0..t}) =  \left(\prod^t_{s=1}\frac{\exp\left[\beta u^{\rm sym}\left(\theta(s)\right)\sigma(s)\right]}{2\cosh\left[\beta u^{\rm sym}\left(\theta(s)\right)\right]  }\right)p_0\left(\sigma^0\right)\label{eq:PPAsymAnsatz}\:, \\ 
\fl  P^{\rm d}(\sigma^{0..t}) =  \left(\prod^t_{s=1}\frac{\exp\left[\beta u^{\rm{d}}\: \sigma(s)\right]}{2\cosh\left[\beta u^{\rm{d}}\right]  }\right)p_0(\sigma^0)\label{eq:PPADAnsatz}\:, \\ 
\fl  P^{\rm real}(\sigma^{0..t}) =  \left(\prod^t_{s=1}\frac{\exp\left[\beta u\: \sigma(s)\right]}{2\cosh\left[\beta u\right]  }\right)p_0(\sigma^0)\label{eq:PPADRAnsatz}\:.
\end{eqnarray}
We insert (\ref{eq:PPAsymAnsatz}), (\ref{eq:PPADAnsatz}) and (\ref{eq:PPADRAnsatz}) in the update functions $\mathcal{F}$ given by equations (\ref{eq:FD}), (\ref{eq:FSym}) and (\ref{eq:FReal}) and close the equations through (\ref{eq:closed}).   
We define the function $u=\mathcal{U}_{k^{\rm in}, k^{\rm sym}}\left(\left\{u_\ell, J_\ell\right\}; \theta\right)$ through the explicit solution $u$ of the implicit equation
\begin{eqnarray}
\fl u =\frac{1}{2\beta}\sum_{\sigma}\sigma\log\left\{\sum_{\btau, \tau}
\frac{\exp\left[\beta \sigma (\sum^{k^{\rm in}}_{\ell=1} J_\ell \tau_\ell + \theta)\right]}{2\cosh\left[\beta \left(\sum^{k^{\rm in}}_{\ell=1} J_\ell \tau_\ell + \theta\right)\right]}\right.
\nonumber \\ \left.
\frac{\exp\left[\beta\sum^{k^{\rm sym}}_{\ell=1} \tau_\ell u_\ell\left(J_\ell \tau\right) + \beta\sum^{k^{\rm in}}_{\ell=k^{\rm sym}+1} \tau_\ell u_\ell \right]}{\prod^{k^{\rm sym}}_{\ell=1}\cosh\left[\beta u_\ell(J_\ell\tau)\right]} \exp\left[\beta \tau u\right]\right\}  \:. \label{eq:u}
\end{eqnarray}
In the sequel we use bimodal distributions of the form (\ref{eq:distriRho}).  We then have two type of messages propagating along symmetric links: $u(\theta) = u(J_0 \sigma) = u^{\sigma}$. 
Using the one-time approximation in (\ref{eq:distriout}) we get for the density of the fields propagating along directed edges
\begin{eqnarray}
  \fl W^{\rm{d}}(u)= \sum^{\infty}_{k^{\rm out}\geq 0}\sum^{\infty}_{k^{\rm in}\geq 0}\sum^{\rm{min}\it (k^{\rm out}, k^{\rm in})}_{k^{\rm sym}=0}\frac{p(k^{\rm in}, k^{\rm out},k^{\rm sym})(k^{\rm out}-k^{\rm sym})}{c_{\rm{out}}-c_{\rm sym}}
\nonumber \\ 
\fl 
\prod^{k^{\rm sym}}_{\ell=1} \int dJ_{\ell}R(J_{\ell})\int du^{+}_\ell du^{-}_\ell \: W^{\rm sym}(u^{+}_\ell, u^{-}_\ell)\prod^{k^{\rm in}}_{\ell = k^{\rm sym} + 1}\int dJ_{\ell}R(J_{\ell})\int du_\ell \: W^{\rm{d}}\left(u_\ell\right) 
\nonumber \\ 
 \delta \left[u - \mathcal{U}_{k^{\rm in}, k^{\rm sym}}\left(\left\{u_{\ell'}, J_{\ell'}\right\}_{\ell'=1..k^{\rm in}}; 0\right)\right] \label{eq:oneTimeWD}\:.
\end{eqnarray}
The density of the fields propagating along the symmetric edges follows from the equations (\ref{eq:distrisym}) 
\begin{eqnarray}
  \fl W^{\rm sym}(u^+, u^-)=  \sum^{\infty}_{k^{\rm in}\geq 0}\sum^{k^{\rm in}}_{k^{\rm sym}=0}\frac{p(k^{\rm in}, k^{\rm sym})k^{\rm sym}}{c_{\rm sym}} \prod^{k_{\rm{in}-1}}_{\ell = k^{\rm sym}}\int dJ_{\ell}R(J_{\ell})\int du_\ell \: W^{\rm{d}}\left(u_\ell\right) 
\nonumber \\ 
\prod^{k^{\rm sym}-1}_{\ell = 1} \int dJ_{\ell}R(J_{\ell})\int du^{+}_\ell du^{-}_\ell \: W^{\rm sym}(u^{+}_\ell, u^{-}_\ell)
\nonumber \\ 
\delta \left[u^- - \mathcal{U}_{k^{\rm in}-1, k^{\rm sym}-1}\left(\left\{u_{\ell'}, J_{\ell'}\right\}_{\ell'=1..k^{\rm in}-1}; -1\right)\right]
\nonumber \\
 \delta\left[u^+ - \mathcal{U}_{k^{\rm in}-1, k^{\rm sym}-1}\left(\left\{u_{\ell'}, J_{\ell'}\right\}_{\ell'=1..k^{\rm in}-1}; 1\right)\right]\:. \label{eq:oneTimeWS}
\end{eqnarray}

Substitution of the one-time approximation (\ref{eq:PPADRAnsatz}) in the equation for the distribution of the real marginals (\ref{eq:distrireal}) gives
\begin{eqnarray}
 \fl W^{\rm real}(u) = \sum^{\infty}_{k^{\rm in}\geq 0}\sum^{k^{\rm in}}_{u=0}p(k^{\rm in}, k^{\rm sym})\prod^{k^{\rm sym}}_{\ell=1} \int dJ_{\ell}R(J_{\ell})\int du^{+}_\ell du^{-}_\ell \: W^{\rm sym}(u^{+}_\ell, u^{-}_\ell)
\nonumber \\ 
\fl 
\prod^{k^{\rm in}}_{\ell = k^{\rm sym}+1}\int dJ_{\ell}R(J_{\ell})\int du_\ell \: W^{\rm{d}}\left(u_\ell\right) \delta\left[u - \mathcal{U}_{k^{\rm in}, k^{\rm sym}}\left(\left\{u_{\ell'}, J_{\ell'}\right\}_{\ell'=1..k^{\rm in}}; 0\right)\right] \label{eq:oneTimeWR}
\end{eqnarray}
Because of the analogy with the equations (\ref{eq:WU}) we call the equations (\ref{eq:oneTimeWD}) and (\ref{eq:oneTimeWS}) the density evolution equations in the one-time approximation.  We see that now we have two densities instead of one: the density for the fields propagating along symmetric edges and the density for the fields propagating along directed edges.  The equation for the marginals on the original graph (\ref{eq:oneTimeWR}) is the equivalent of (\ref{eq:WR}) for the symmetric model.  Another important difference is that the update function $\mathcal{U}_{k^{\rm in}, k^{\rm sym}}$, for partially asymmetric graphs, is not explicitly known.  Instead we have to solve the implicit equation (\ref{eq:u}).  The equations (\ref{eq:oneTimeWD}), (\ref{eq:oneTimeWS}) and (\ref{eq:oneTimeWR}) constitute an algorithm that generalizes belief propagation to graphs with asymmetric bonds. 

\subsection{Fully Symmetric and Fully Asymmetric Cases}
For graphs with exclusively symmetric or asymmetric links, we determine the explicit form of $\mathcal{U}_{k^{\rm in}, k^{\rm sym}}$.
For a fully symmetric graph, the equation (\ref{eq:u}) admits the solution
\begin{eqnarray}
 \fl \mathcal{U}_{k^{\rm in}-1, k^{\rm sym}-1}\left(\left\{u_\ell, J_\ell\right\}_{\ell=1..k^{\rm sym}-1}; \theta\right) = \theta + \beta^{-1}\sum^{k^{\rm sym}-1}_{\ell=1}\atanh\left[\tanh \left(\beta J_\ell\right) \tanh \left(\beta u_\ell\right)\right] \:,\label{eq:SymSol}
\end{eqnarray}
where we used $u_\ell(J_\ell \tau) = J_\ell \tau + u_\ell$.  For fully asymmetric models we have the solution
\begin{eqnarray}
 \fl \mathcal{U}_{k^{\rm in}, k^{\rm sym}}\left(\left\{u_\ell, J_\ell\right\}_{\ell=1..k^{\rm in}}; 0\right) =  \frac{1}{2\beta}\sum_{\sigma}\sigma\log\left\{\sum_{\btau, \tau}
\frac{\exp\left[\beta \sigma \sum_\ell J_\ell \tau_\ell \right]}{2\cosh\left[\beta \left(\sum_\ell J_\ell \tau_\ell \right)\right]} \exp\left[\beta\sum^{k^{\rm in}}_{\ell=1} \tau_\ell u_\ell\right] 
\right\} \:. \nonumber \\\label{eq:AsymSol}
\end{eqnarray}
From (\ref{eq:SymSol}) and (\ref{eq:AsymSol}) it follows that the one-time approximation gives the correct results for Glauber dynamics of the Ising model on fully symmetric and asymmetric graphs.  Now we check how good the equations (\ref{eq:oneTimeWD}), (\ref{eq:oneTimeWS}) and (\ref{eq:oneTimeWR}) are for models on partially asymmetric graphs.  
\subsection{Results and Comparison with Simulations}
\begin{figure}[h!]
\begin{center}
\includegraphics[angle=-90, width=.6 \textwidth]{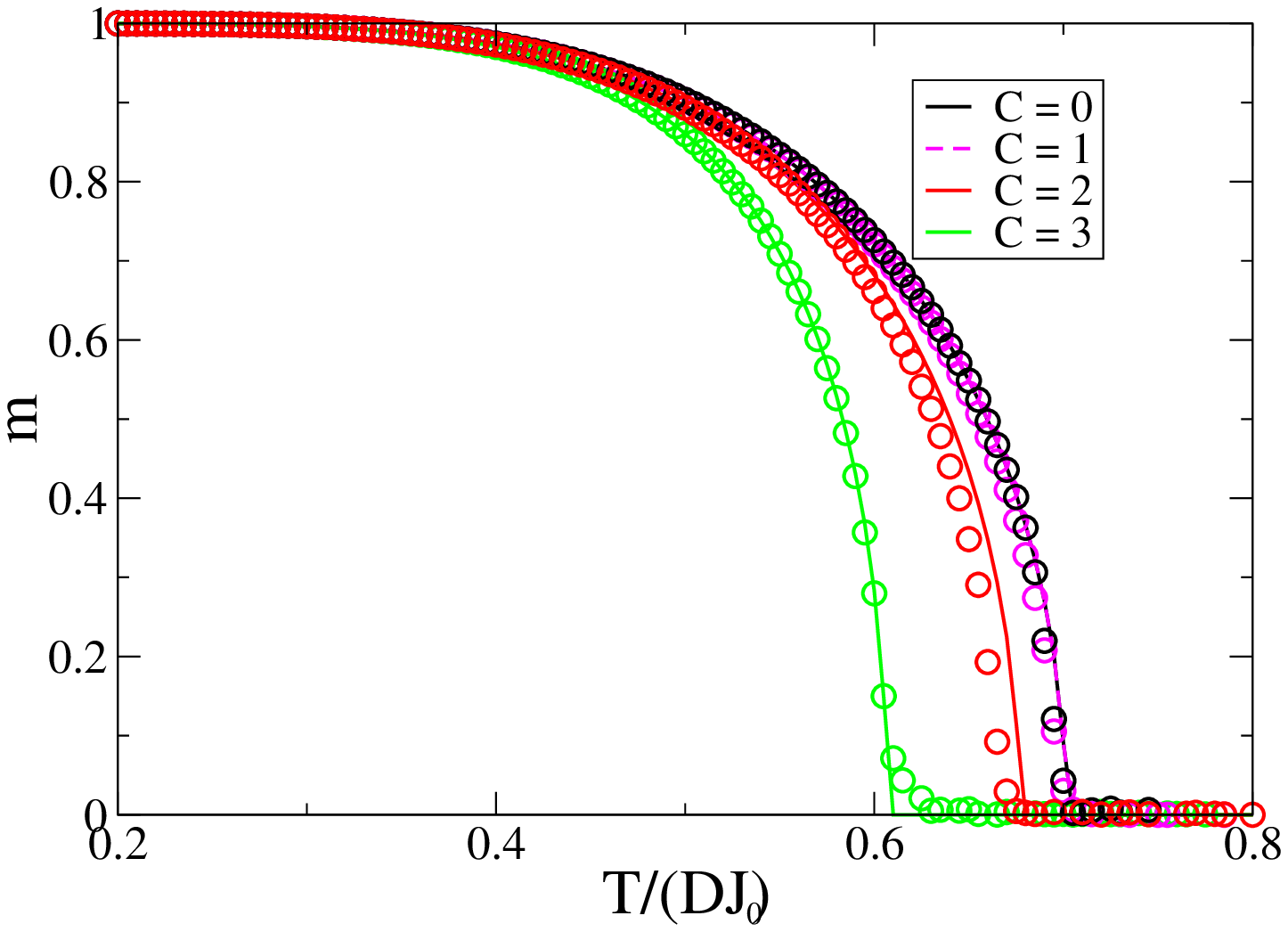}
\caption{The magnetisation $m$ as a function of the rescaled temperature $T/(DJ_0)$ for a ferromagnet (i.e. $\rho=1$) defined on a  Bethe lattice.  The indegree $D$ of the graph equals $3$ and each site is incident to $C$ symmetric bonds. The simulations (markers) are mean values of $20$ runs on graphs of sizes $\mathcal{O}(10^4)$.  The theory (lines) follow from solving recursively the density evolution equations in the one-time approximation.} \label{fig:C3rho1}
\end{center}
\begin{center}
\includegraphics[angle=-90, width=.6 \textwidth]{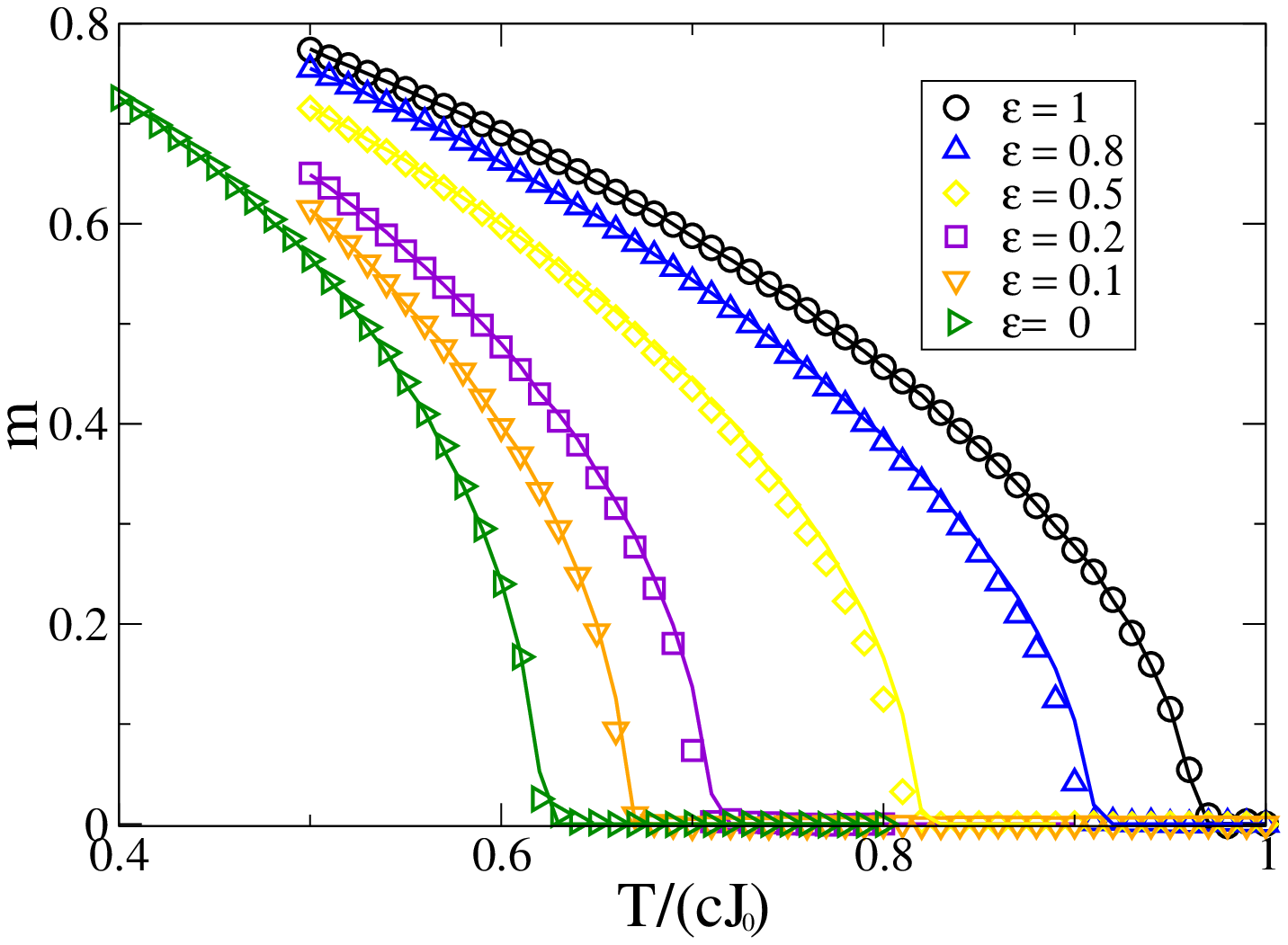}
\caption{The magnetisation $m$ as a function of the temperature $T/(cJ_0)$ for an Ising model without bond disorder on a Poissonian graph with mean connectivity $c=3$ and different fractions of symmetric edges $\epsilon$.  The lines are obtained by population dynamics from the density evolution equations in the one-time approximation for populations of sizes $\mathcal{O}(10^5)$.  The markers are the average results from $20$ runs with the heat-bath algorithm on a graph instance of size $\mathcal{O}(10^5)$.} \label{fig:poisson}
\end{center}
\end{figure}

\begin{figure}[h!]
 \begin{center}
\includegraphics[angle=-90, width=.6 \textwidth]{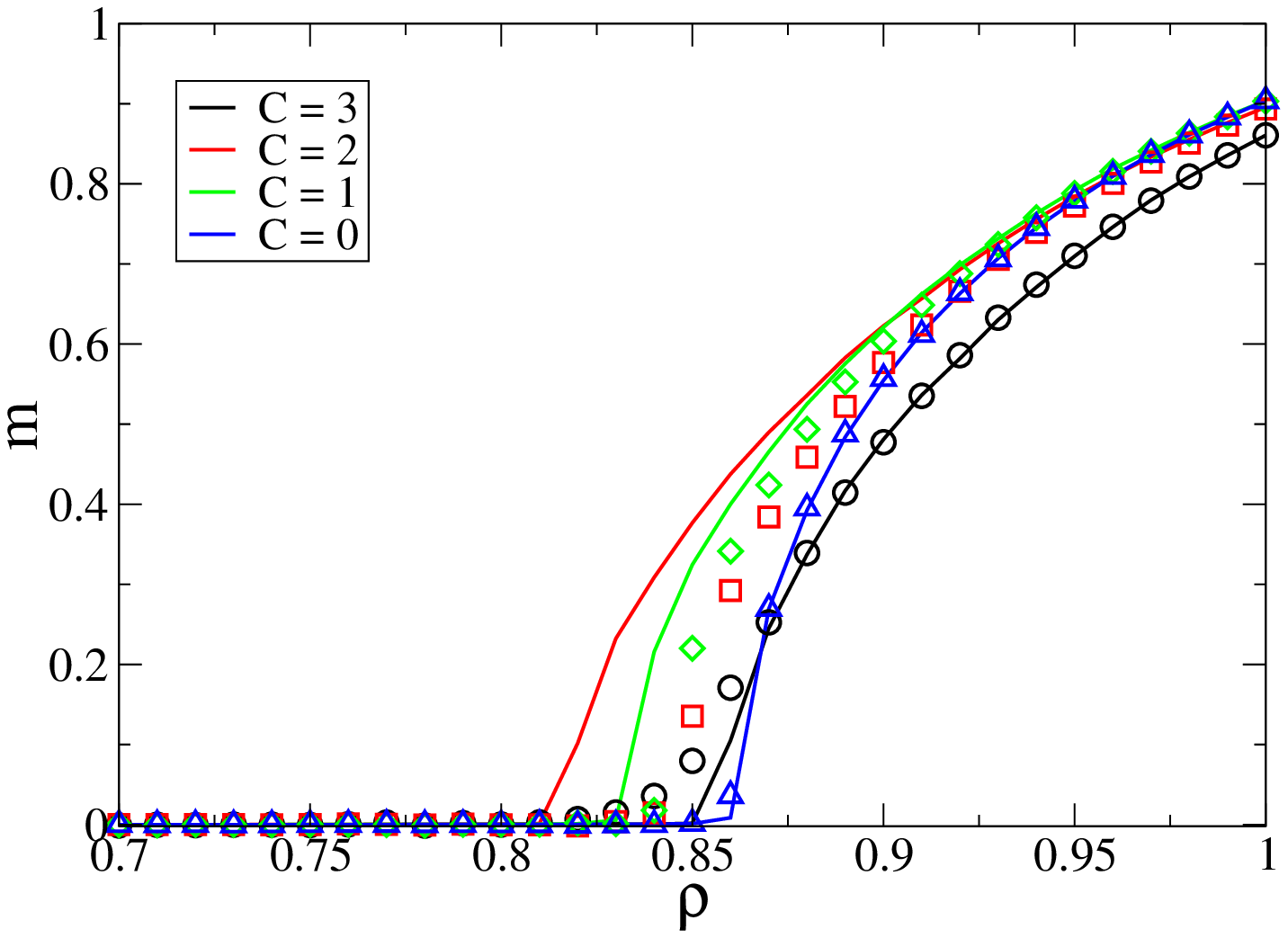}
\caption{The magnetisation $m$ as a function of the bias $\rho$ in the couplings  (see equation (\ref{eq:distriRho})) at a temperature $T/(D J_0)=0.5$.  The Ising model is defined on a Bethe lattice with the degree distribution defined in (\ref{eq:perfBethe}).  The indegree $D$ equals $3$ and results are shown for various values of the symmetric degree $C$. The simulations (markers) are the mean values of $20$ runs on graphs of sizes $\mathcal{O}(10^5)$.  The theory (lines) are results from the  density evolution equations in the one-time approximation using a Monte Carlo calculation with populations of $\mathcal{O}(10^5)$ fields.} \label{fig:T0.5}
\end{center}
\end{figure}

We numerically solve the equations   (\ref{eq:oneTimeWD}), (\ref{eq:oneTimeWS}) and (\ref{eq:oneTimeWR}) with a Monte Carlo integration.  The unknown distributions $W^{\rm d}$, $W^{\rm sym}$ and $W^{\rm real}$ are represented as populations.  The procedure is also known as population dynamics \cite{Mez2001}. 
In figure \ref{fig:C3rho1} the magnetisation is plotted as a function of the temperature for an Ising model without bond disorder on a Bethe lattice. The degree distribution is then: 
\begin{eqnarray}
p(k^{\rm in}, k^{\rm out},k^{\rm sym}) = \delta(k^{\rm sym} -C)\delta(k^{\rm in}-D)p(k^{\rm out})\:. \label{eq:perfBethe}
\end{eqnarray}
 Because there is no disorder the distributions $W^{\rm{d}}(u)$ and $W^{\rm sym}(u^+, u^-)$ are delta functions.  Figure \ref{fig:C3rho1} tells us that the theory and the simulations are in good agreement. For $C=0, 3$ the results coincide while for $C=1, 2$ there is a small deviation.      In figure \ref{fig:poisson} we checked how good the approximation is on graphs with fluctuating connectivities.  We simulated an Ising model without bond disorder on Poissonian graphs drawn from the ensemble defined in \ref{app:PoisDegreeI}.  Despite the fluctuations in the degrees we find a good agreement between theory and simulations. This confirms that for models without bond disorder, the one-time approximation works very well.
In figure \ref{fig:T0.5} we plot the magnetisation on a Bethe lattice with bond disorder. We see that the difference between theory and simulation increases with the bias $\rho$ in the bonds. 
   
\subsection{Bifurcation Analysis}
We determine the P to F and the P to SG transition lines for Ising models with bimodal distributions using a bifurcation analysis around the paramagnetic solution. First we note that the equations (\ref{eq:oneTimeWD}) and  (\ref{eq:oneTimeWS}) admit the solution:
\begin{eqnarray}
 W^{\rm{d}}(u) = \delta(u) \:,\label{eq:P3}\\ 
W^{\rm{s}}(u^+, u^-) = \int dA \:W^P(A)\delta(u^+-A)\delta(u^-+A) \:. \label{eq:P4}
\end{eqnarray}
Indeed, when we insert (\ref{eq:P3}) and (\ref{eq:P4}) in (\ref{eq:oneTimeWS}) we get for $W^P(A)$: 
\begin{eqnarray}
\fl W^P(A) = \sum^{\infty}_{k^{\rm in}\geq 0}\sum^{k^{\rm in}}_{k^{\rm sym}=0}\frac{p(k^{\rm in}, k^{\rm sym})k^{\rm sym}}{c_{\rm sym}} \prod_{0< \ell\leq k^{\rm sym}-1} \int dA_\ell W^P(A_\ell)
\nonumber \\  
\delta\left[A-\mathcal{A}_{k^{\rm sym}-1, k^{\rm in}-1}\left( \left\{A_{\ell'}\right\}_{\ell'=1..k^{\rm sym}-1}\right)\right]
\end{eqnarray}
with $\mathcal{A}_{k^{\rm sym}-1, k^{\rm in}-1}$ the explicit solution of 
\begin{eqnarray}
\fl  A =
\frac{1}{2\beta}\sum_{\theta = \pm1}\theta\log\left\{\sum_{\btau}
\frac{\exp\left[\theta \beta J_0(\sum^{k^{\rm in}-1}_{\ell=1} \tau_\ell + 1)\right]}{\cosh\left[\beta J_0\left(\sum^{k^{\rm in}-1}_{\ell=1}  \tau_\ell + 1\right)\right]}
\cosh\left[\beta\sum^{k^{\rm sym}-1}_{\ell=1}  \tau_\ell A_\ell  + \beta  A \right]\right\} \:. \nonumber \\ \label{eq:AExpl}
\end{eqnarray}
The solution (\ref{eq:P3}) and (\ref{eq:P4}) represents the paramagnetic phase. 
For  a symmetric lattice with $k^{\rm in} = k^{\rm sym}$ we have $\mathcal{A}_{k^{\rm sym}-1, k^{\rm in}-1}= J_0$.   For $k^{\rm sym} = 0$ we obtain
\begin{eqnarray}
\fl  \mathcal{A}_{k^{\rm sym}-1, k^{\rm in}-1} = \frac{1}{2\beta}\sum_{\theta=\pm1}\theta\log\left\{\sum_{\btau}
\frac{\exp\left[\theta \beta J_0(\sum^{k^{\rm in}-1}_{\ell=1} \tau_\ell + 1)\right]}{\cosh\left[\beta J_0\left(\sum^{k^{\rm in}-1}_{\ell=1}  \tau_\ell + 1\right)\right]}\right\} \:.
\end{eqnarray}
For partially asymmetric graphs $W^P(A)$ results in a more complicated form.

To calculate the P to F and P to SG transitions, we expand the equations (\ref{eq:oneTimeWD}),  (\ref{eq:oneTimeWS}) and (\ref{eq:oneTimeWR}) around the solution given by (\ref{eq:P3}) and (\ref{eq:P4}). We use an expansion in $\alpha \ll 1$ with:
\begin{eqnarray}
 \int W^{\rm{d}}(u) u^n = m^{\rm{d}}_n \sim \mathcal{O}(\alpha^n) \label{eq:exp0}\:,
\end{eqnarray}
and 
\begin{eqnarray}
\fl m^{\rm{s}}_n(\sigma) = \int du^+du^- W(u^+, u^-)\left(u(\sigma)- \sigma A\right)^n \sim \mathcal{O}(\alpha^n) \label{eq:exp1}\:, \\ 
\fl m^{s}_{nm}  = \int du^+du^- W(u^+, u^-) \left(u(+)-A\right)^n \left(u(-)+A\right)^m \sim  \mathcal{O}(\alpha^{n+m}) \label{eq:exp2}\:.
\end{eqnarray}
Substitution of (\ref{eq:exp0}), (\ref{eq:exp1}) and (\ref{eq:exp2}) in  (\ref{eq:oneTimeWD}),  (\ref{eq:oneTimeWS}) and (\ref{eq:oneTimeWR}) gives, up to linear order in $\alpha$, the equation
\begin{eqnarray}
 \left(\begin{array}{c} m^{\rm{d}}_1\\ m^{\rm{s}}_1 \end{array}\right) = \rho \bM \left(\begin{array}{c} m^{\rm{d}}_1 \\ m^{\rm{s}}_1 \end{array}\right)\:, \label{eq:bif10}
\end{eqnarray}
with 
\begin{eqnarray}
\bM = \left(\begin{array}{cc} M_{\rm{dd}}&M_{\rm{ds}} \\ M_{\rm{sd}}&M_{\rm{ss}} \end{array}\right) \:.
\end{eqnarray}
The critical temperature $T_F$ of the P to F phase transition is given by the equation $|\lambda|^{-1}(T_F) = \rho$ with $\lambda(T)$ the eigenvalue of $\bM$ with the largest modulus.
\begin{figure}[h!]
\begin{center}
\includegraphics[angle=-90, width=.6 \textwidth]{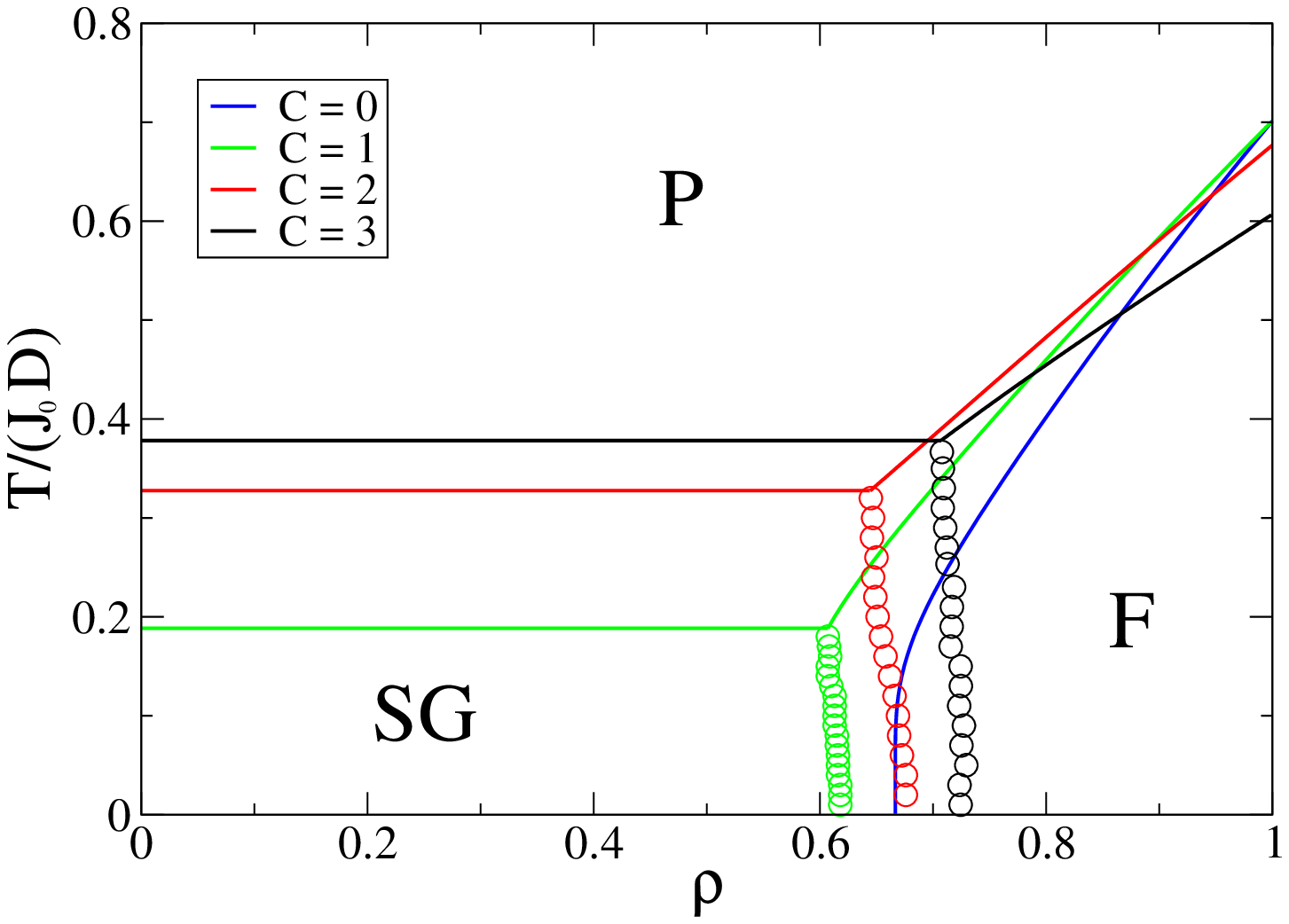}
\caption{The phase transition lines between the P-phase, F-phase and SG-phase on a Bethe lattice with $k^{\rm in}=3$ and $k^{\rm sym} = C$.  The lines are calculated from the equations for the critical temperatures derived from a bifurcation analysis.  The markers are computed with the population dynamics method on the density evolution equations in the one-time approximation.} \label{fig:BetheAsSym2}
\end{center}
\end{figure}
The elements of the $\bM$ matrix are
\begin{eqnarray}
\fl M_{\rm{dd}}=  \sum^{\infty}_{k^{\rm out}\geq 0}\sum^{\infty}_{k^{\rm in}\geq 0}\sum^{\rm{min}\it(k^{\rm out}, k^{\rm in})}_{k^{\rm sym}=0}\frac{p(k^{\rm in}, k^{\rm out},k^{\rm sym})(k^{\rm out}-k^{\rm sym})}{c_{\rm{out}}-c_{\rm sym}} \left(k^{\rm in}-k^{\rm sym}\right) \:,
\nonumber \\
 \int \prod^{k^{\rm sym}}_{\ell=1}dA_\ell W(A_\ell) \left[\frac{ \langle \tau_{k^{\rm sym}+1}\rangle_d}{1 - \langle \tau \rangle_{d}}\right]\:,\\ 
\fl M_{\rm{ds}} = \sum^{\infty}_{k^{\rm out}\geq 0}\sum^{\infty}_{k^{\rm in}\geq 0}\sum^{\rm{min}\it(k^{\rm out}, k^{\rm in})}_{k^{\rm sym}=0}\frac{p(k^{\rm in}, k^{\rm out},k^{\rm sym})(k^{\rm out}-k^{\rm sym})}{c_{\rm{out}}-c_{\rm sym}} k^{\rm sym}
\nonumber \\
 \int \prod^{k^{\rm sym}}_{\ell=1}dA_\ell W(A_\ell) \left[\frac{ \langle \tau_1\rangle_d -   \tanh\left(\beta A_1\right)\langle \tau\rangle_d }{1 - \langle \tau \rangle_{d}}\right]\:,\\
\fl M_{\rm{ss}} = \sum^{\infty}_{k^{\rm in}\geq 0}\sum^{k^{\rm in}}_{k^{\rm sym}=0}\frac{p(k^{\rm in}, k^{\rm sym})k^{\rm sym}}{c_{\rm sym}}(k^{\rm sym}-1)
\nonumber \\
\int \prod^{k^{\rm sym}-1}_{\ell=1}dA_\ell W(A_\ell)\frac{\langle \tau_1\rangle_s - \tanh\left(\beta A_1\right)\langle \tau \rangle_s}{1-\langle \tau\rangle_s}\:,\\ 
\fl M_{\rm{sd}} = \sum^{\infty}_{k^{\rm in}\geq 0}\sum^{k^{\rm in}}_{k^{\rm sym}=0}\frac{p(k^{\rm in}, k^{\rm sym})k^{\rm sym}}{c_{\rm sym}} (k^{\rm in}-k^{\rm sym}) \int \prod^{k^{\rm sym}-1}_{\ell=1}dA_\ell W(A_\ell) \frac{\langle  \tau_{k^{\rm sym}}\rangle_s}{1-\langle \tau\rangle_s} \:.
\end{eqnarray}
The averages $\langle \cdot \rangle_d$ and $\langle \cdot \rangle_s$ appearing in the matrix elements are given by
\begin{eqnarray}
\fl  \langle  f(\btau, \tau, \bA)\rangle_d =  \sum_{\btau, \tau} \mathcal{W}_{k^{\rm sym}, k^{\rm in}}\left(1, \tau, \btau;0, \mathcal{A}\left( \left\{A_\ell\right\}_{1, k^{\rm sym}}\right)\right)f(\btau, \tau, \bA)\:, 
\end{eqnarray}
and 
\begin{eqnarray}
\fl \langle  f(\btau, \tau, \bA)\rangle_s=
2^{-1} \sum_{\btau, \tau, \sigma}\sigma\mathcal{W}_{k^{\rm sym}-1, k^{\rm in}-1}\left( \sigma, \tau, \btau; J_0, \mathcal{A}\left( \left\{A_\ell\right\}_{1, k^{\rm sym}-1}\right)\right) f(\btau, \tau, \bA)\:, \nonumber \\ 
\end{eqnarray}
in the weight $\mathcal{W}_{k^{\rm sym}, k^{\rm in}}$. 
The weight $\mathcal{W}_{k^{\rm sym}, k^{\rm in}}$ is expressed as
\begin{eqnarray}
\fl \mathcal{W}_{k^{\rm sym}, k^{\rm in}}(\sigma, \tau, \btau;\theta, A)\equiv \frac{\frac{\exp\left[\beta \sigma\left(J_0 \sum^{k^{\rm in}}_{\ell=1} \tau_\ell + \theta\right)\right]}{2\cosh\left[\beta \left(J_0 \sum^{k^{\rm in}}_{\ell=1} \tau_\ell + \theta\right)\right]} \exp\left[\beta \tau \left(\sum^{k^{\rm sym}}_{\ell=1}  A_\ell \tau_\ell + \rm{sign}\left(\theta\right)\it A\right)\right]}{\sum_{\btau, \tau}\frac{\exp\left[\beta \sigma\left(J_0 \sum^{k^{\rm in}}_{\ell=1} \tau_\ell + \theta\right)\right]}{2\cosh\left[\beta \left(J_0 \sum^{k^{\rm in}}_{\ell=1} \tau_\ell + \theta\right)\right]}\exp\left[\beta \tau \left(\sum^{k^{\rm sym}}_{\ell=1}  A_\ell \tau_\ell + \rm{sign}\left(\theta\right)\it A\right)\right]}\:. 
\nonumber \\ 
\end{eqnarray}
For symmetric graphs the bifurcation condition $|\lambda|^{-1} = \rho$ of (\ref{eq:bif10}) simplifies to $m^{\rm{s}}_1 = \rho M_{\rm{ss}}m^{\rm{s}}_1$, which leads to the condition
\begin{eqnarray}
  \rho \tanh\left(\beta_F J_0\right) \sum^{\infty}_{k^{\rm in}\geq 0} p(k^{\rm sym})\frac{k^{\rm sym}(k^{\rm sym}-1)}{c^{\rm sym}} = 1\:. \label{eq:TF}
\end{eqnarray}
Equation (\ref{eq:TF}) gives the P to F transition line on a symmetric graph \cite{Hartmann2005}.  For fully asymmetric graphs we get $m^{\rm{d}}_1 = \rho M_{\rm{dd}}m^{\rm{d}}_1$: 
\begin{eqnarray}
 \fl m^{\rm{d}}_1 =  \rho m^{\rm{d}}_1\sum^{\infty}_{k^{\rm in}\geq 0}p(k^{\rm in})2^{-k^{\rm in}} \sum_{\btau}\left(\sum^{k^{\rm in}}_{\ell=1}\tau_\ell\right)  \tanh\left(\beta_F J_0\sum^{k^{\rm in}}_{\ell=1} \tau_\ell\right)\:.
\end{eqnarray}
This is the same condition we found in (\ref{eq:TransFerro}). 

To determine the P to SG transition we expand up to order $\alpha^2$ with $\int du \: W^{\rm{d}}(u) u = 0$ and  $\int du\: W^{\rm{s}}(u^+, u^-) (u(\sigma)-\sigma A) = 0$.  We then arrive at
\begin{eqnarray}
 \left(\begin{array}{c} m^{\rm{d}}_2\\ m^{\rm{s}}_2 \end{array}\right) = \bQ \left(\begin{array}{c} m^{\rm{d}}_2 \\ m^{\rm{s}}_2 \end{array}\right)\:, \  \bQ = \left(\begin{array}{cc} Q_{\rm{dd}}&Q_{\rm{ds}} \\ Q_{\rm{sd}}&Q_{\rm{ss}} \end{array}\right) \:, \label{bif:SG}
\end{eqnarray}
with
\begin{eqnarray}
\fl Q_{\rm{dd}}=  \sum^{\infty}_{k^{\rm out}\geq 0}\sum^{\infty}_{k^{\rm in}\geq 0}\sum^{\rm{min}\it(k^{\rm out}, k^{\rm in})}_{k^{\rm sym}=0}\frac{p(k^{\rm in}, k^{\rm out},k^{\rm sym})(k^{\rm out}-k^{\rm sym})}{c_{\rm{out}}-c_{\rm sym}} \left(k^{\rm in}-k^{\rm sym}\right) 
\nonumber \\
\int \prod^{k^{\rm sym}}_{\ell=1}dA_\ell W(A_\ell) \left[\frac{ \langle \tau_{k^{\rm sym}+1}\rangle_d}{1 - \langle \tau \rangle_{d}}\right]^2\:,\\ 
\fl Q_{\rm{ds}} = \sum^{\infty}_{k^{\rm out}\geq 0}\sum^{\infty}_{k^{\rm in}\geq 0}\sum^{\rm{min}\it (k^{\rm out}, k^{\rm in})}_{k^{\rm sym}=0}\frac{p(k^{\rm in}, k^{\rm out},k^{\rm sym})(k^{\rm out}-k^{\rm sym})}{c_{\rm{out}}-c_{\rm sym}} k^{\rm sym}
\nonumber \\
\int \prod^{k^{\rm sym}}_{\ell=1}dA_\ell W(A_\ell) \left[\frac{ \langle \tau_1\rangle_d -   \tanh\left(\beta A_1\right)\langle \tau\rangle_d }{1 - \langle \tau \rangle_{d}}\right]^2\:,\\
\fl Q_{\rm{ss}} = \sum^{\infty}_{k^{\rm in}\geq 0}\sum^{k^{\rm in}}_{k^{\rm sym}=0}\frac{p(k^{\rm in}, k^{\rm sym})k^{\rm sym}}{c_{\rm sym}}(k^{\rm sym}-1) 
\nonumber \\
 \int \prod^{k^{\rm sym}-1}_{\ell=1}dA_\ell W(A_\ell)\left[\frac{\langle \tau_1\rangle_s - \tanh\left(\beta A_1\right)\langle \tau \rangle_s}{1-\langle \tau\rangle_s}\right]^2\:,\\ 
\fl Q_{\rm{sd}} = \sum^{\infty}_{k^{\rm in}\geq 0}\sum^{k^{\rm in}}_{k^{\rm sym}=0}\frac{p(k^{\rm in}, k^{\rm sym})k^{\rm sym}}{c_{\rm sym}} (k^{\rm in}-k^{\rm sym}) \int \prod^{k^{\rm sym}-1}_{\ell=1}dA_\ell W(A_\ell) \left[\frac{\langle  \tau_{k^{\rm sym}}\rangle_s}{1-\langle \tau\rangle_s}\right]^2 \:.
\end{eqnarray}
\begin{figure}[h!]
\begin{center}
\includegraphics[angle=-90, width=.6 \textwidth]{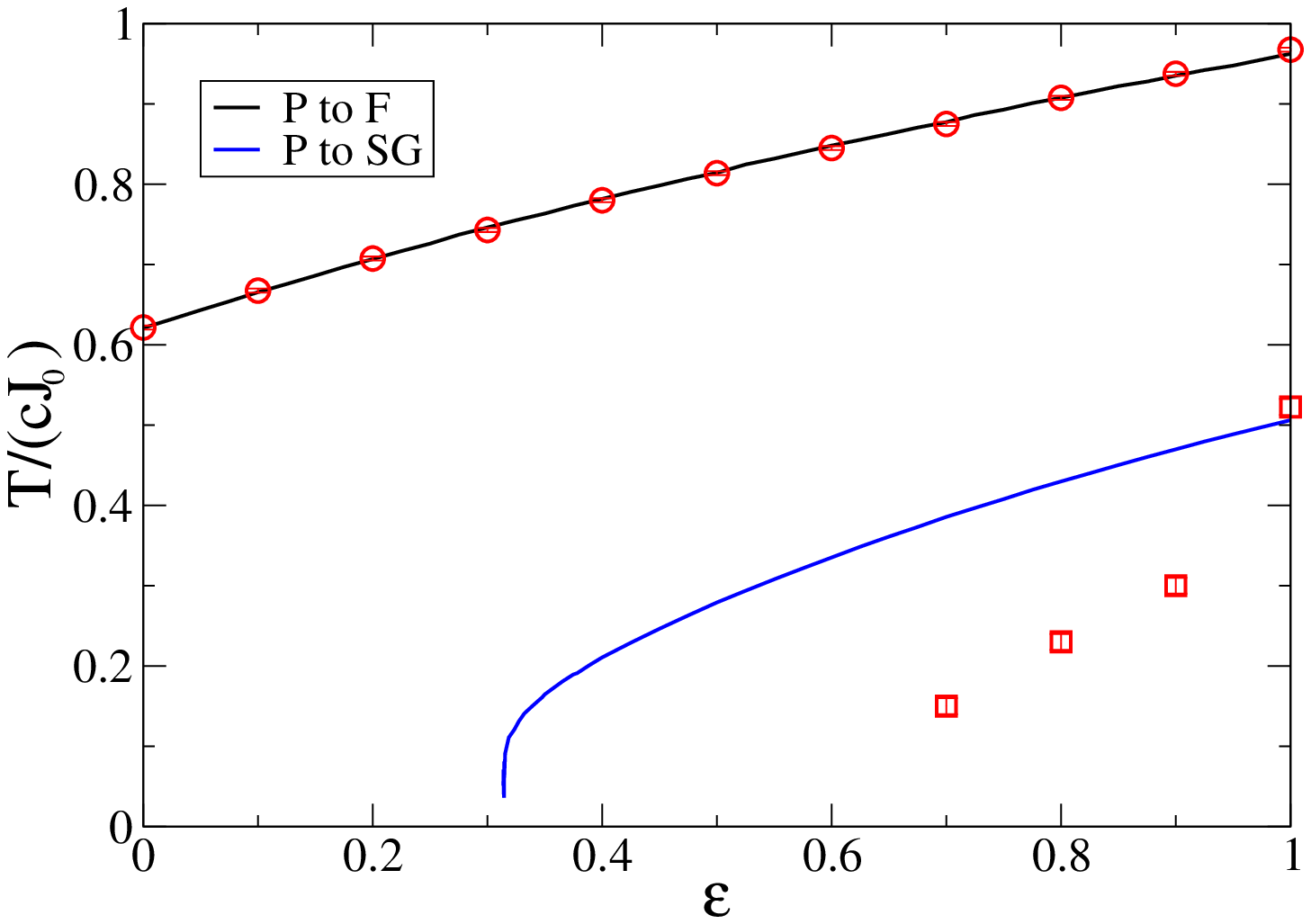}
\caption{Critical temperatures computed from a bifurcation analysis (lines) within the one-time approximation are compared with simulations (markers) as a function of the fraction of symmetric edges $\epsilon$.  The results shown for the P to F transition are for an Ising model on a Poissonian graph without bond disorder.  The results shown for the P to SG transition are for an Ising model on a Poissonian graph for $\rho=0$.} \label{fig:PoissonAsSym2}
\end{center}
\end{figure}
Again we find the correct bifurcation condition 
\begin{eqnarray}
  \tanh^2\left(\beta_{SG} J_0\right) \sum^{\infty}_{k^{\rm in}\geq 0} p(k^{\rm sym})\frac{k^{\rm sym}(k^{\rm sym}-1)}{c^{\rm sym}} = 1\:,
\end{eqnarray}
for symmetric lattices \cite{Hartmann2005}.  
In figure \ref{fig:BetheAsSym2} the phase transition lines between the P-phase, F-phase and SG-phase are shown for a bond-disordered Ising model with a bimodal distribution.  The P to F and P to SG lines are obtained using the bifurcation analysis while the F to SG transition is found through population dynamics.  In figure \ref{fig:PoissonAsSym2} we present the P to F transition line for a ferromagnet ($\rho=1$) and the P to SG transition line obtained through the bifurcation analysis on a Poissonian graph as a function of the fraction of symmetric edges $\epsilon$.  The markers from simulations are in agreement with the theory.  The simulations are performed through the method of Binder cumulants, see \cite{Binder2005}.  In \ref{sec:simulations} we give some details on how we derived these critical temperatures.  We see that the SG-phase dissappears when the asymmetry in the graph increases.  

\section{Conclusion}\label{sec:disc}
In this paper we applied the cavity method to study the dynamics of spin models on a given graph instance. We derived a set of effective equations which describe the dynamics.  Solving these recursive equations can be seen as the equivalent of the belief propagation algorithm known from inference problems or decoding algorithms.  Just like the latter, we expect these equations to be exact on a tree.  The main difference with statics is that path probabilities, instead of stationary probabilities of single spins, are propagated along the edges of the graph.  We took the average over an ensemble of graphs to find the recursive equations describing the dynamics of Ising models on typical graphs drawn from this ensemble.  These equations generalize the result of \cite{Hat2005} to graphs with arbitrary degree distributions. The macroscopic evolution of the system is given as a function of three mean values of path probabilities: the one of the probabilities propagating along directed edges, the one of the probabilities propagating along symmetric edges and the one of the marginal path probabilities of the spins on the original graph.  On a Poissonian graph the three recursive equations for these path probabilities reduce to one and we find back the result derived in \cite{Hat2005}.   We solved the following problems on the basis of these equations.

The evolution of macroscopic observables like the magnetisation is calculated for a small number of time steps because the computational complexity of the derived equations scales exponentially in time.  These results are compared with simulations and both methods are in good agreement.   This confirms that the derived equations are exact on graphs with a local tree structure.

The effective equations simplify on fully asymmetric graphs.  The effective dynamics becomes Markovian and the stationary distribution is derived.  We studied the phase diagrams of an Ising model on a Bethe lattice showing the existence of a paramagnetic, ferromagnetic and a chaotic phase.  We also examined a neural network with Hebbian couplings on a scale-free graph where we found that correlations between indegrees and outdegrees increase the retrieval properties of the neural network tremendously.  

We introduced an heuristic method to derive the stationary distribution of Ising like models on partially asymmetric graphs by assuming that
the path probabilities factorize in time.  The equations are closed such that we recover the correct stationary distribution for models on fully symmetric and fully asymmetric graphs.  This method is approximative.  We found a set of equations that generalize belief propagation to an algorithm that computes the marginals of the stationary distribution of Ising models on graphs with asymmetric couplings.  The evolution of macroscopic observables and the phase transitions are obtained with this method and compared with simulations.  Without bond disorder the theory and the simulations give almost identical results. When introducing bond disorder the theoretical predictions start to fail and a more refined approximation is necessary in that case.  

 In future work we will try to get better algorithms that derive the marginals of the stationary solution by using a systematic method that minimizes a functional like the Kullback-Leibler distance.  Work along this way is in progress. Furthermore, it would be interesting to analyze the dynamics of a neural network, for example given in \cite{Morita2001}, with the cavity equations.

\appendix

\section{The Poissonian Ensemble}\label{app:PoisDegreeI}
The Poissonian ensemble $\mathcal{G}_{\rm{p}}(c, \epsilon)$ is defined through the following probability function
\begin{eqnarray}
 P_{\rm p}(\bC; c,\epsilon) \equiv \prod_{i<j}\left(\frac{c}{N}\delta(c_{ij}; 1) + \left(1-\frac{c}{N}\right)\delta(c_{ij}; 0) \right) 
\nonumber \\ 
 \prod_{i>j}\left(\epsilon \delta(c_{ij}; c_{ji}) + \left(1-\epsilon\right)\left(\frac{c}{N}\delta(c_{ij}; 1) + \left(1-\frac{c}{N}\right)\delta(c_{ij}; 0) \right) \right)\:. \label{eq:EnsemblePoiss1}
\end{eqnarray}
The mean connectivity is given by $c$ while $\epsilon$ is the fraction of symmetric edges.  This is the same ensemble as used in \cite{Hat2004}.  In the recursive equations (\ref{eq:Pout}), (\ref{eq:PSym}) and (\ref{eq:PReal}) for dynamics we need to know the degree distribution.  
 We calculate the degree distribution $p_{\rm{p}}$, defined through
\begin{eqnarray}
 \fl p_{\rm p}(k^{\rm in}, k^{\rm out}, k^{\rm sym}) \equiv \frac{\sum_{i}\sum_{\bC}P(\bC)\delta(k^{\rm in}; \sum_{j}c_{ij})\delta(k^{\rm out};\sum_{j}c_{ji})\delta\left(k^{\rm sym};\sum_{j}c_{ij}c_{ji}\right)}{N} \:.  \label{eq:degrees}
\end{eqnarray}
We write the Kronecker $\delta$ as
\begin{eqnarray}
 \delta(k^{\rm in}; \sum_{j}c_{ij}) = \int^{2\pi}_0 \frac{d\omega}{2\pi} \: \exp\left[i\omega(\sum_{j}c_{ij} -k^{\rm in} )\right] \:,\\
 \delta(k^{\rm out}; \sum_{j}c_{ji}) = \int^{2\pi}_{0} \frac{d\omega'}{2\pi}\: \exp\left[i\omega'(\sum_{j}c_{ji} -k^{\rm out} )\right]\:, \\ 
\delta\left(k^{\rm sym}; \sum_{j}c_{ij}c_{ji}\right) = \int^{2\pi}_0 \frac{d\zeta}{2\pi}  \: \exp\left[i\zeta(\sum_{j}c_{ji}c_{ij} -k^{\rm sym})\right] \:.
\end{eqnarray}
Substitution of the above integrals in (\ref{eq:degrees}) leads to
\begin{eqnarray}
\fl p_{\rm p}(k^{\rm in}, k^{\rm out}, k^{\rm sym}) = \int \frac{d\omega}{2\pi} \frac{d\omega'}{2\pi} \frac{d\zeta}{2\pi} \:  \exp\left[-i\omega k^{\rm in} - i\omega' k^{\rm out} -  i\zeta k^{\rm sym}\right] \nonumber \\ 
\fl  \left(\epsilon \frac{c}{N}e^{\omega + \omega' + \zeta}  + \epsilon \left(1-\frac{c}{N}\right) + (1-\epsilon)\frac{c}{N}\left(1-\frac{c}{N}\right)\left(e^{\omega} + e^{\omega'}\right) + (1-\epsilon)\left(1-\frac{c}{N}\right)^2\right)^{N-1}\:.  \nonumber 
\end{eqnarray}
We keep only the $\mathcal{O}\left(N^0\right)$ terms:
\begin{eqnarray}
 \fl p_{\rm p}(k^{\rm in}, k^{\rm out}, k^{\rm sym})  =  \int \frac{d\omega}{2\pi} \frac{d\omega'}{2\pi} \frac{d\zeta}{2\pi} \exp\left[-i\omega k^{\rm in} - i\omega' k^{\rm out} - i\zeta k^{\rm sym}\right]
\nonumber \\ 
\exp\left[c\epsilon\left( e^{\omega+\omega'+\zeta} - 1\right)+ \left(1-\epsilon\right)c\left(e^{\omega}+e^{\omega'}\right) - 2\left(1-\epsilon\right)c\right]\:. \label{eq:last}
\end{eqnarray}
When we expand the exponentials in (\ref{eq:last}) we get
\begin{eqnarray}
 \fl p_{\rm p}(k^{\rm in}, k^{\rm out}, k^{\rm sym})  =   \int \frac{d\omega}{2\pi} \frac{d\omega'}{2\pi} \frac{d\zeta}{2\pi} \exp\left[-i\omega k^{\rm in} - i\omega' k^{\rm out} - i\zeta k^{\rm sym}  -c\epsilon -2\left(1-\epsilon\right)c\right]
\nonumber \\ 
\fl \sum_{n_1}\left(c\epsilon\right)^{n_1}\frac{\exp\left[i n_1\left(\omega+\omega'+\zeta\right)\right]}{n_1!}\sum_{n_2}\left(c(1-\epsilon)\right)^{n_2}\frac{\exp\left[i n_2\omega \right]}{n_2!}\sum_{n_3}\left(c(1-\epsilon)\right)^{n_3}\frac{\exp\left[i n_3\omega'\right]}{n_3!} \:. \nonumber 
\end{eqnarray}
Integrating over $\zeta$ gives us
\begin{eqnarray}
 \fl p_{\rm p}(k^{\rm in}, k^{\rm out}, k^{\rm sym})  =p_{\rm{p}}\left(k^{\rm sym}; c\epsilon\right) \int \frac{d\omega}{2\pi} \frac{d\omega'}{2\pi}  \exp\left[-i \omega k^{\rm in} - i \omega' k^{\rm out} -2\left(1-\epsilon\right)c\right]
\nonumber \\ 
\fl  \sum_{n_2}\left(c(1-\epsilon)\right)^{n_2}\frac{\exp\left[i(n_2+k^{\rm sym})\omega \right]}{n_2!}\sum_{n_3}\left(c(1-\epsilon)\right)^{n_3}\frac{\exp\left[i(n_3+k^{\rm sym})\omega'\right]}{n_3!} \:, \nonumber 
\end{eqnarray}
with $p_{\rm{p}}(k;c) = \exp(-c)c^k/k!$.
Finally we integrate over $\omega$ and $\omega'$:
\begin{eqnarray}
  \fl p_{\rm p}(k^{\rm in}, k^{\rm out}, k^{\rm sym}) 
 = p_{\rm{p}}\left(k^{\rm sym}; c\epsilon\right)p_{\rm{p}}\left(k^{\rm in}-k^{\rm sym}; c(1-\epsilon)\right)p_{\rm{p}}\left(k^{\rm out}-k^{\rm sym}; c(1-\epsilon)\right)\:. \nonumber \\ 
\end{eqnarray}
 When we sum out the outdegrees we finally obtain
\begin{eqnarray}
 \fl p_{\rm p}(k^{\rm in}, k^{\rm sym}) =\frac{e^{-c}c^{k^{\rm in}}}{k^{\rm in}!}\left(\begin{array}{c} k^{\rm in}\\ k^{\rm sym} \end{array}\right)\epsilon^{k^{\rm sym}}\left(1-\epsilon\right)^{k^{\rm in}-k^{\rm sym}}\:. \label{eq:degPoiss1B}
\end{eqnarray}
For the Poissonian ensembles the equations for the dynamics (\ref{eq:Pout}), (\ref{eq:PSym}) and (\ref{eq:PReal}) simplify because of the property
\begin{eqnarray}
\fl  \sum_{k^{\rm in}, k^{\rm sym}} p_{\rm p}(k^{\rm in}, k^{\rm sym}) \frac{k^{\rm sym}}{c_{\rm sym}} A\left(k^{\rm sym}, k^{\rm in}\right)=  \sum_{k^{\rm in}, k^{\rm sym}} p_{\rm p}(k^{\rm in}, k^{\rm sym}) A\left(k^{\rm sym}+1, k^{\rm in}+1\right) \:.
\nonumber \\ \label{eq:change}
\end{eqnarray}
Indeed, repeatedly applying (\ref{eq:change}) on (\ref{eq:Pout}), (\ref{eq:PSym}) and (\ref{eq:PReal}) gives $\overline{P}^{\rm real} = \overline{P}^{\rm sym}$ and when $\theta^{1..t} = 0^{1..t}$: $\overline{P}^{\rm{d}} = \overline{P}^{\rm sym}$. The three equations reduce to one equation.

\section{Stationary State for Glauber Dynamics}\label{sec:equil}
By using the tools of equilibrium statistical mechanics we determine the stationary behaviour of a model evolving through Glauber dynamics determined by the equations (\ref{eq:W}) and (\ref{eq:GlauberSigma}). We consider the case where the field equals
\begin{eqnarray}
 h_i(s) = h_i(\bsigma(s-1)) = \sum_{j}J_{ij}c_{ij}\sigma_j(s-1) \:,
\end{eqnarray}
with $J_{ij} = J_{ji}$ and $c_{ij} = c_{ji}$.  The stationary state is given by $P_{\rm{st}}(\bsigma) \sim \exp\left(-\beta H(\bsigma)\right)$, with $H(\bsigma)$ equal to 
\begin{eqnarray}
 H(\bsigma) = -\frac{1}{\beta}\sum^N_{i=1}\log\left(2\cosh\left(\beta h_i(\bsigma)\right)\right) \:. \label{eq:HamEquil}
\end{eqnarray}
This system has the same thermodynamic behaviour as a model with the two-spin Hamiltonian
\begin{eqnarray}
 H(\bsigma, \btau) = -\sum_{i,j}J_{ij}c_{ij}\sigma_i\tau_j \:.
\end{eqnarray}

Performing the standard replica method or the cavity method (see for example \cite{Nikos2003}) we get the following set of selfconsistent equations for the distribution $W$ of the cavity fields $u$, $v$ and $w$, within the replica symmetric approximation: 
 \begin{eqnarray}
\fl  W^c(u, v,w) = \sum^{\infty}_{k=0}  \frac{p(k)k}{c} \int \prod^{k-1}_{\ell=1}\left(dJ_\ell R(J_{\ell})\right) \int \prod^{k-1}_{\ell=1}\left(du_{\ell} dv_\ell dw_\ell W^c(u_{\ell}, v_\ell,w_\ell) \right) 
\nonumber \\ 
 \delta\left[u- \sum^{k-1}_{\ell=1}\mathcal{U}(J_\ell, u_{\ell}, v_\ell,w_\ell)\right] \delta\left[v- \sum^{k-1}_{\ell=1}\mathcal{V}(J_\ell, u_{\ell}, v_\ell,w_\ell)\right]
\nonumber \\
 \delta\left[w- \sum^{k-1}_{\ell=1}\mathcal{W}(J_\ell, u_{\ell}, v_\ell,w_\ell)\right] \:, \label{eq:WSymEquil}
\end{eqnarray}
with 
\begin{eqnarray}
\fl  \mathcal{U}(J, u, v,w) = \frac{1}{4\beta}\sum_{s,t}s\log\left\{\sum_{\tilde{s}, \tilde{t}}\exp\left[\beta \left( J s\tilde{t} + J\tilde{s}t + u  \tilde{s} + v \tilde{t} + w \tilde{s}\tilde{t}\right) \right]\right\} \:, \label{eq:UCalc} \\
 \fl \mathcal{V}(J, u, v,w) = \frac{1}{4\beta}\sum_{s,t}t\log\left\{\sum_{\tilde{s}, \tilde{t}}\exp\left[\beta \left( J s\tilde{t} + J\tilde{s}t + u  \tilde{s} + v \tilde{t} + w \tilde{s}\tilde{t}\right) \right]\right\}\:, \label{eq:VCalc} \\
 \fl \mathcal{W}(J, u, v,w) = \frac{1}{4\beta}\sum_{s,t}st\log\left\{\sum_{\tilde{s}, \tilde{t}}\exp\left[\beta \left( J s\tilde{t} + J\tilde{s}t + u  \tilde{s} + v \tilde{t} + w \tilde{s}\tilde{t}\right) \right]\right\} \:. \label{eq:WCalc}
\end{eqnarray}
The distribution $W^c$ determines the density of the fields $(u^{(j)}_i, v^{(j)}_i, w^{(j)}_i)$ along the edges of the graph.
 These fields correspond with the marginal probability distribution $P^{(j)}_i$ on the cavity graph $G^{(j)}$:  
\begin{eqnarray}
 P^{(j)}_i(\sigma, \tau) = \frac{\exp\left[u^{(j)}_i \sigma + v^{(j)}_i \tau + w^{(j)}_i \sigma \tau\right]}{\sum_{\sigma, \tau}\exp\left[u^{(j)}_i \sigma + v^{(j)}_i \tau + w^{(j)}_i \sigma \tau\right]} \:.
\end{eqnarray}

On a given graph instance it is possible to calculate the fields $(u^{(j)}_i, v^{(j)}_i, w^{(j)}_i)$ through the belief propagation algorithm. In the $n$-th time step of the algorithm, messages $(u^{(n)}_{i\rightarrow j},v^{(n)}_{i\rightarrow j},w^{(n)}_{i\rightarrow j})$ are propagated along the edges of the graph.  The update equations of these messages are given by
\begin{eqnarray}
u^{(n+1)}_{i\rightarrow j} = \sum_{\ell \in \partial_i\setminus j} \mathcal{U}(J_{\ell i}, u^{(n)}_{\ell \rightarrow i}, v^{(n)}_{\ell \rightarrow i}, w^{(n)}_{\ell \rightarrow i})\:, \label{eq:cav1}\\
v^{(n+1)}_{i\rightarrow j} = \sum_{\ell \in \partial_i\setminus j} \mathcal{V}(J_{\ell i}, u^{(n)}_{\ell \rightarrow i}, v^{(n)}_{\ell \rightarrow i}, w^{(n)}_{\ell \rightarrow i})\:, \label{eq:cav2}\\
w^{(n+1)}_{i\rightarrow j} = \sum_{\ell \in \partial_i\setminus j} \mathcal{W}(J_{\ell i}, u^{(n)}_{\ell \rightarrow i}, v^{(n)}_{\ell \rightarrow i}, w^{(n)}_{\ell \rightarrow i})\:. \label{eq:cav3}
\end{eqnarray}
The fields $(u^{(j)}_i, v^{(j)}_i, w^{(j)}_i)$ are given by the solutions of (\ref{eq:cav1}), (\ref{eq:cav2}) and (\ref{eq:cav3}) for $n \rightarrow \infty$. We call the equations (\ref{eq:cav1}), (\ref{eq:cav2}) and (\ref{eq:cav3}) the cavity or belief propagation equations.
The density of the messages $(u^{(n)}_{i\rightarrow j},v^{(n)}_{i\rightarrow j},w^{(n)}_{i\rightarrow j})$ can be found in equation (\ref{eq:WSymEquil}).  That is why (\ref{eq:WSymEquil}) is also called the density evolution equation. 

When we choose $W^c(u,v,w) = \delta(w)\delta(u-v)W^c(u)$ we find 
\begin{eqnarray}
  \fl W^c(u) = \sum^{\infty}_{k=0}  \frac{p(k)k}{c} \int \prod^{k-1}_{\ell=1}\left(dJ_\ell R(J_{\ell})\right) \int \prod^{k-1}_{\ell=1}\left(du_{\ell}W^c(u_{\ell}) \right) 
\nonumber \\
 \delta \left[ u -\beta^{-1}\sum^{k-1}_{\ell=1}\atanh\left(\tanh(\beta J_\ell)\tanh(\beta u_\ell)\right)\right]\:. \label{eq:WU}
\end{eqnarray}
The distribution $W^r$  of the real fields $u$ of the single site marginals  $\sum_{\bsigma\setminus \sigma_i}P_{\rm st}(\bsigma) \sim \exp\left[u_i\sigma\right]$ are given by the selfconsistent equation
\begin{eqnarray}
 \fl  W^r(u) = \sum^{\infty}_{k=0}  \frac{p(k)k}{c} \int \prod^{k}_{\ell=1}\left(dJ_\ell R(J_{\ell})\right) \int \prod^{k}_{\ell=1}\left(du_{\ell}W^c(u_{\ell}) \right) 
\nonumber \\
 \delta \left[ u -\beta^{-1}\sum^{k}_{\ell=1}\atanh\left(\tanh(\beta J_\ell)\tanh(\beta u_\ell)\right)\right]\:. \label{eq:WR}
\end{eqnarray}
The solution $W(u,v,w) = \delta(w)\delta(u-v)W^c(u)$ corresponds with the solution $w^{(n+1)}_{i\rightarrow j} = 0$ and $u^{(n+1)}_{i\rightarrow j} = v^{(n+1)}_{i\rightarrow j}$ of the equations (\ref{eq:cav1}), (\ref{eq:cav2}) and (\ref{eq:cav3}).  The cavity equations  reduce in this case to
\begin{eqnarray}
 u^{(n+1)}_{i\rightarrow j} = \beta^{-1}\sum_{\ell \in \partial_i\setminus j} \atanh\left(\tanh\left(\beta J_{\ell i}\right) \tanh\left(\beta u^{(n)}_{\ell \rightarrow i}\right)\right)\:. \label{eq:CavFinal}
\end{eqnarray}
The single site marginals $P_i(\sigma)$ can be calculated through
\begin{eqnarray}
 u_i = \beta^{-1}\lim_{n\rightarrow \infty}\sum_{\ell \in \partial_i} \atanh\left(\tanh\left(\beta J_{\ell i}\right) \tanh\left(\beta u^{(n)}_{\ell \rightarrow i}\right)\right)\:. \label{eq:CavMarg} 
\end{eqnarray}
\section{Simulations}\label{sec:simulations}
\begin{figure}[h!]
\begin{center}
\includegraphics[angle=-90, width=.6 \textwidth]{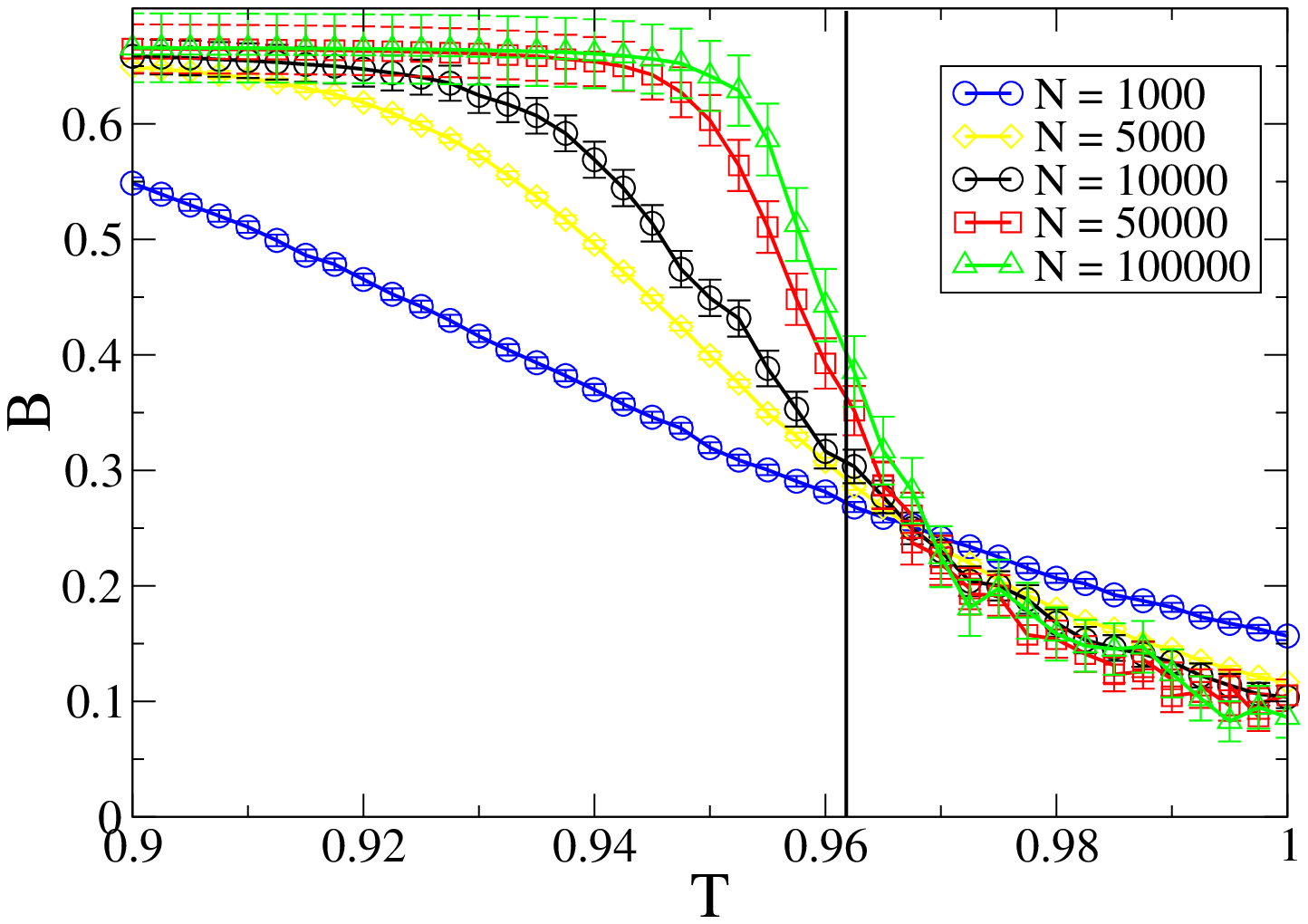}
\caption{The Binder cumulant $B_N$ for the P to F transition on a Poissonian graph with $\epsilon=1$, $c = 3$ and different system sizes $N$.  The vertical line indicates the critical value $T_F = \left(c \: \atanh(1/c)\right)^{-1}$.  We see that the Binder cumulant method overestimates slightly the critical value for small system sizes.} \label{fig:PoissonBinder}
\end{center}
\end{figure}

To determine the critical temperature from the P-phase to the F-phase of Ising models on a graph of size $N$ we calculate the Binder cumulant $B_N$ \cite{Binder2005}: 
\begin{eqnarray}
 B_N(T) \equiv 1 - \frac{1}{3} \frac{\langle m^4\rangle}{\langle m^2\rangle^2}\:.
\end{eqnarray}
The brackets denote an average over the stationary distribution.  The quantity $m$ is the magnetisation $\sum_i \sigma_i/N$.  
In the F-phase $B=2/3$ while in the P-phase $B = 0$.  The point $T_F$ where the $B_N(T)$ lines for different values of $N$ cross is the critical temperature.  In figure \ref{fig:PoissonBinder} we present $B_N$ as a function of $T$ for a Poissonian graph with $\epsilon=1$ and $c = 3$.  The value $T_F$ for finite system sizes is found to be higher then the theoretical result $T_F =  \left(c \: \atanh(1/c)\right)^{-1}$ valid for $N\rightarrow \infty$.  

The critical temperature for the P-SG transition is calculated through $A_N$ \cite{Bhatt}:
\begin{eqnarray}
 A_N(T) \equiv \overline{\frac{1}{2}\left(3-\frac{\langle q^4\rangle}{\langle q^2\rangle^2}\right)}
\end{eqnarray}
The bar denotes the average over the quenched variables.  
Above the critical temperature $T_{SG}$ the lines for different system sizes join into one line.  

\section*{References}
\bibliographystyle{ieeetr} 
\bibliography{bibliography}

\begin{thebibliography}{10}

\bibitem{Corr2006}
L.~Correale, M.~Leone, A.~Pagnani, M.~Weigt, and R.~Zecchina, ``The
  computational core and fixed point organization in boolean networks,'' {\em
  J. Stat. Mech.: Theory Exp.}, vol.~3, p.~03002, 2006.

\bibitem{Coolen2005}
A.~C.~C. Coolen, {\em The Mathematical Theory of Minority Games: Statistical
  Mechanics of Interacting Agents}.
\newblock Oxford University Press, USA, 2005.

\bibitem{Paczus2000}
M.~Paczuski, K.~E. Bassler, and A.~Corral, ``Self-organized networks of
  competing boolean agents,'' {\em Phys. Rev. Lett.}, vol.~84, p.~2185, 2000.

\bibitem{Vic1999}
R.~Vicente, D.~Saad, and Y.~Kabashima, ``Finite-connectivity systems as
  error-correcting codes,'' {\em Phys. Rev. E}, vol.~60, p.~5352, 1999.

\bibitem{Klein07}
J.~Kleinberg, {\em Cascading Behavior in Networks: Algorithmic and Economic
  Issues}, ch.~24, pp.~613--633.
\newblock Cambridge University Press, 2007.

\bibitem{viana1985}
L.~Viana and A.~J. Bray, ``Phase diagrams for dilute spin glasses,'' {\em J.
  Phys. C: Solid State Phys}, vol.~18, p.~3037, 1985.

\bibitem{Mez2001}
M.~M\'ezard and G.~Parisi, ``The bethe lattice spin glass revisited,'' {\em
  Eur. Phys. J. B}, vol.~20, p.~217, 2001.

\bibitem{Kschi2001}
F.~R. Kschischang, B.~J. Frey, and H.~Loeliger, ``Factor graphs and the
  sum-product algorithm,'' {\em IEEE Transactions on Information Theory},
  vol.~47, pp.~498--519, 2001.

\bibitem{Flor2008}
F.~Krzakala, A.~Montanari, F.~Ricc-Tersenghi, G.~Semerjian, and L.~Zdeborova,
  ``Gibbs states and the set of solutions of random constraint satisfaction
  problems,'' {\em Proc. Natl. Acad. Sci.}, vol.~104, p.~10318, 2007.

\bibitem{Tim2008}
T.~Rogers, I.~P. Castillo, R.~K\"uhn, and K.~Takeda, ``Cavity approach to the
  spectral density of sparse symmetric random matrices,'' {\em Phys. Rev. E.},
  vol.~78, p.~031116, 2008.

\bibitem{Steiner}
M.~Bayati, C.~Borgs, A.~Braunstein, J.~Chayes, A.~Ramezanpour, and R.~Zecchina,
  ``Statistical mechanics of steiner trees,'' {\em Phys. Rev. Lett.}, vol.~101,
  p.~037208, 2008.

\bibitem{Cris1988}
A.~Crisanti and H.~Sompolinsky, ``Dynamics of spin systems with randomly
  asymmetric bonds: Ising spins and glauber dynamics,'' {\em Phys. Rev. A},
  vol.~12, pp.~4865--4874, 1988.

\bibitem{Dominicis1978}
C.~D. Dominicis, ``Dynamics as a substitute for replicas in systems with
  quenched random impurities,'' {\em Phys. Rev. B}, vol.~18, pp.~4913--4919,
  1978.

\bibitem{Mimura}
K.~Mimura and A.~C.~C. Coolen, ``Parallel dynamics of disordered ising spin
  systems on finitely connected random graphs with arbitrary degree
  distributions.'' in preparation.

\bibitem{Hat2005}
J.~P.~L. Hatchett, I.~P. Castillo, A.~C.~C. Coolen, and N.~S. Skantzos,
  ``Dynamical replica analysis of disordered ising spin systems on finitely
  connected random graph,'' {\em Phys. Rev. Lett.}, vol.~95, p.~117204, 2005.

\bibitem{Mimura2}
K.~Mimura and A.~C.~C. Coolen, ``Generating functional analysi of ldgm channel
  coding with many short loops.'' to be published in Proceedings of the 2009
  IEEE International Symposium on Information Theory (ISIT2009, Seoul, Korea),
  509-513.

\bibitem{Mozeika2007}
A.~Mozeika and A.~C.~C. Coolen, ``Dynamical replica analysis of processes on
  finitely connected random graphs i: vertex covering,'' {\em J. Phys. A: Math.
  Theor.}, vol.~41, p.~115003.

\bibitem{Kiemes2008}
M.~Kiemes and H.~Horner, ``Dynamics of an ising spin glass on the bethe
  lattice,'' {\em J. Phys. A: Math. Gen.}, vol.~41, p.~324017, 2008.

\bibitem{Binder2005}
K.~Binder and D.~W. Heermann, {\em Monte Carlo Simulations in Statistical
  Physics}.
\newblock Springer, 2002.

\bibitem{Hat2004}
J.~P.~L. Hatchett, B.~Wemmenhove, I.~P. Castillo, T.~Nikoletopoulos, N.~S.
  Skantzos, and A.~C.~C. Coolen, ``Parallel dynamics of disordered ising spin
  systems on finitely connected random graphs,'' {\em J Phys A: Math Gen},
  vol.~37, p.~6201, 2004.

\bibitem{Der1987}
B.~Derrida, ``Dynamical phase transition in non-symmetric spin glasses,'' {\em
  J Phys A: Math Gen}, vol.~20, pp.~L721--L725, 1987.

\bibitem{Der1987Co}
B.~Derrida, E.~Gardner, and A.~Zippelius, ``An exactly solvable asymmetric
  neural network model,'' {\em Europhysics Letters}, vol.~4, pp.~167--173,
  1987.

\bibitem{Eiss1992}
H.~Eissfeller and M.~Opper, ``New method for studying the dynamics of
  disordered spin systems without finite-size effects,'' {\em Phys. Rev.
  Lett.}, vol.~68, p.~2094, 1992.

\bibitem{Victor2005}
V.~M. Egu\'iluz, D.~R. Chialvo, G.~A. Cecchi, M.~Baliki, and A.~V. Apkarian,
  ``Scale-free brain functional networks,'' {\em Phys. Rev. Lett.}, vol.~94,
  p.~018102, 2005.

\bibitem{Isaac2004}
I.~P. Castillo, B.~Wemmenhove, J.~P.~L. Hatchett, A.~C.~C. Coolen, N.~S.
  Skantzos, and T.~Nikoletopoulos, ``Analytic solution of attractor neural
  networks on scale-free graphs,'' {\em J. Phys. A: Math. Gen.}, vol.~37,
  p.~8789, 2004.

\bibitem{Schwart2002}
N.~Schwartz, R.~Cohen, D.~ben Avraham, A.~L. Barab\'asi, and S.~Havlin,
  ``Percolation in directed scale-free networks,'' {\em Phys. Rev. E},
  p.~015104, 2002.

\bibitem{Bor2000}
J.~M. Borwein, D.~M. Bradley, and R.~E. Crandall, ``Computational strategies
  for the riemann zeta function,'' {\em J. Comput. Appl. Math.}, vol.~121,
  pp.~247--296.

\bibitem{Hartmann2005}
A.~K. Hartmann and M.~Weigt, {\em Phase Transitions in Combinatorial
  Optimization Problems: Basics, Algorithms and Statistical Mechanics}, ch.~5
  Statistical mechanics of the Ising model.
\newblock Wiley-VCH, 2005.

\bibitem{Morita2001}
S.~Morita, K.~Oshio, Y.~Osana, Y.~Funabashi, K.~Oka, and K.~Kawamura,
  ``Geometrical structure of the neuronal network of caenorhabditis elegans,''
  {\em Physica A}, vol.~298, p.~553, 2001.

\bibitem{Nikos2003}
I.~P. Castillo and N.~S. Skantzos, ``The little-hopfield model on a random
  graph,'' {\em J Phys A}, vol.~37, p.~9087, 2004.

\bibitem{Bhatt}
R.~N. Bhatt and A.~P. Young, ``Numerical studies of ising spin glasses in two,
  three, and four dimensions,'' {\em Phys. Rev. B}, vol.~37, p.~5606, 1988.

\end{thebibliography}

\end{document}